\documentclass[a4paper]{article}

\usepackage{amsmath, amsfonts, amsthm}
\usepackage{subcaption}
\usepackage{tikz}
\usepackage[round]{natbib}
\usepackage{enumitem}
\usepackage[margin=1.2in]{geometry}
\usepackage[hidelinks]{hyperref}

\hypersetup{
  pdftitle={Adaptively-structured mixed models for simple clustered data},
  pdfauthor={Helen Ogden},
  pdfkeywords={Longitudinal data; Mixed-effects models; Penalized likelihood; Smoothing}
}

\providecommand{\T}{\mathrm{T}}

\providecommand{\alttext}[1]{}
\renewcommand{\alttext}[1]{}

\newtheorem{theorem}{Theorem}
\newtheorem{lemma}{Lemma}

\usetikzlibrary{shapes.geometric}
\usetikzlibrary{shapes.misc}

\graphicspath{{figures/}}

\definecolor{colAdaStruMM}{HTML}{D55E00}
\definecolor{colLocalGAM}{HTML}{E69F00}
\definecolor{colHGAM}{HTML}{009E73}
\definecolor{colPACE}{HTML}{0072B2}
\definecolor{colFACE}{HTML}{CC79A7}
\definecolor{colBayesFPCA}{HTML}{56B4E9}
\definecolor{colSITAR}{HTML}{666666}

\DeclareRobustCommand{\legAdaStruMM}{\raisebox{1pt}{\tikz{
      \draw[colAdaStruMM,solid,line width=0.9pt](0,0) -- (5mm,0) node
           [midway, draw=colAdaStruMM, fill=colAdaStruMM,
            circle, scale=0.35] {};
}}}

\DeclareRobustCommand{\legLocalGAM}{\raisebox{1.5pt}{\tikz{
      \draw[colLocalGAM,solid,line width=0.9pt](0,0) -- (5mm,0) node
           [midway, draw=colLocalGAM, fill=colLocalGAM,
            regular polygon, regular polygon sides=3, scale=0.25] {};
}}}

\DeclareRobustCommand{\legHGAM}{\raisebox{1pt}{\tikz{
      \draw[colHGAM,solid,line width=0.9pt](0,0) -- (5mm,0) node
           [midway, draw=colHGAM, fill=colHGAM,
            rectangle, scale=0.5] {};
}}}

\DeclareRobustCommand{\legPACE}{\raisebox{1pt}{\tikz{
      \draw[colPACE,solid,line width=0.9pt](0,0) -- (5mm,0) node
           [midway, draw=colPACE, fill=colPACE,
            diamond, aspect=1, scale=0.45] {};
}}}

\DeclareRobustCommand{\legFACE}{\raisebox{-1pt}{\tikz{
      \draw[colFACE,solid,line width=0.9pt](0,0) -- (5mm,0);
      \draw[colFACE,line width=0.7pt] (2.5mm,-1.2mm) -- (2.5mm,1.2mm);
      \draw[colFACE,line width=0.7pt] (1.3mm,0) -- (3.7mm,0);
}}}

\makeatletter
\pgfdeclareshape{crossedbox}{
  \inheritsavedanchors[from=rectangle]
  \inheritanchorborder[from=rectangle]
  \inheritanchor[from=rectangle]{center}
  \backgroundpath{
    \pgfpathrectangle{\pgfpoint{-0.15cm}{-0.15cm}}{\pgfpoint{0.3cm}{0.3cm}}
    \pgfpathmoveto{\pgfpoint{-0.15cm}{-0.15cm}}
    \pgfpathlineto{\pgfpoint{0.15cm}{0.15cm}}
    \pgfpathmoveto{\pgfpoint{-0.15cm}{0.15cm}}
    \pgfpathlineto{\pgfpoint{0.15cm}{-0.15cm}}
  }
}
\makeatother

\DeclareRobustCommand{\legBayesFPCA}{\raisebox{-1pt}{\tikz{
      \draw[colBayesFPCA,solid,line width=0.9pt](0,0) -- (5mm,0) node
           [midway, draw=colBayesFPCA, fill=white,
            shape=crossedbox, scale=0.5, line width=0.8pt] {};
}}}

\DeclareRobustCommand{\legSITAR}{\raisebox{-1pt}{\tikz{
      \draw[colSITAR,solid,line width=0.9pt](0,0) -- (5mm,0);
      \draw[colSITAR,line width=0.8pt] (1.5mm,-1.1mm) -- (3.5mm,1.1mm);
      \draw[colSITAR,line width=0.8pt] (1.5mm,1.1mm) -- (3.5mm,-1.1mm);
}}}

\newcommand{\suppsec}[1]{Appendix~\ref{#1}}
\newcommand{\supplementarymaterialstatement}{}

\title{Adaptively-structured mixed models for simple clustered data}
\author{Helen Ogden\thanks{h.e.ogden@soton.ac.uk}\\
  School of Mathematical Sciences, University of Southampton,\\
  Highfield, Southampton SO17 1BJ, U.K.}
\date{}

\begin{document}

\maketitle

\begin{abstract}
We propose adaptively-structured mixed models for simple clustered
data. Like classical mixed-effects models, they share information
between clusters through random effects, but they estimate the
associated design functions from the data rather than requiring them
to be specified in advance. This retains the mixed-effects mechanism
for information sharing while allowing the structure to adapt flexibly
to the data. We establish consistency and asymptotic normality for
population-level estimation and show that cluster-specific predictions
are asymptotically as accurate as predictions based on the true
population structure. In simulations, adaptively-structured mixed
models substantially improve the quality of inference relative to
existing general-purpose methods while remaining computationally
efficient. An application to body-fat data from adolescent girls
illustrates how the method captures both the average pattern over time
and variation between individuals.

\end{abstract}


\section{Introduction}

In this paper, we propose a new class of flexible mixed-effects models
for simple clustered data. We observe data of the form
\[(x_{ij}, Y_{ij}), \quad i = 1, \ldots, d, \quad j = 1, \ldots, n_i,\]
where $x_{ij} \in \mathbb{R}$ is a covariate and $Y_{ij} \in
\mathbb{R}$ the associated response for the $j$th observation on
cluster $i$. For instance, for simple longitudinal data, clusters
represent individuals and the covariate is time. All models we
consider may be written in the general form
\begin{equation}
  \label{eq:model_Y}
  Y_{ij} = \mu_i(x_{ij}) + \epsilon_{ij}, \quad i = 1, \ldots, d,
  \quad j = 1, \ldots, n_i,
\end{equation}
where $\epsilon_{ij} \sim N(0, \sigma^2)$ is an error term.  We aim to
flexibly model the individual trajectories $\mu_i$.

We have three modelling requirements. We are not aware of any existing
method for simple clustered data which meets all three requirements
simultaneously:
\begin{enumerate}
\item \emph{Individual-trajectory flexibility.} Models should be
  sufficiently flexible to represent individual trajectories as
  arbitrary smooth curves. If there are many observations on cluster
  $i$, we expect estimated trajectories to look like a smooth curve
  fitted to the data from cluster $i$. Classical mixed-effects models
  with pre-specified random-effects designs do not meet this
  requirement in general, because they restrict the possible
  relationships between individual trajectories.
  \label{req:flexible}
\item \emph{Encoding structural similarities.} Models should be able
  to encode structural similarities between clusters. Classical
  mixed-effects models meet this requirement, because the variation in
  trajectories is encoded in a small number of random effects. By
  contrast ``local'' approaches, which fit separate flexible
  regression models for each cluster, do not meet this requirement. As
  a consequence, if there are only a few observations on cluster $i$,
  a local estimate of that trajectory will be highly
  uncertain. \label{req:sharing}
\item \emph{Inferential benefit.} Methods should not only encode
  structural similarities, but also exploit them to improve estimation
  of individual trajectories and population summaries. We will review
  some methods from functional principal components analysis which
  meet requirements \ref{req:flexible} and \ref{req:sharing}, but can
  give less accurate estimates than methods which ignore structural
  similarities, and therefore do not always meet this requirement.
  \label{req:works}
\end{enumerate}

To meet these requirements, we propose Adaptively-Structured Mixed
Models (AdaStruMMs), which replace the pre-specified random-effects
design functions of classical mixed-effects models by unknown
functions estimated from the data. Inference is based on a penalized
likelihood, with the penalty controlling the balance between fit to
the data and smoothness of the trajectories. We establish consistency
and asymptotic normality of the fitted population mean and covariance
functions and show that individual trajectory predictors are
asymptotically equivalent to oracle predictors under the true
population-level parameter. Simulation studies demonstrate that
AdaStruMMs substantially improve the quality of inference relative to
existing general-purpose methods. We focus on a single covariate to
make the methodological development more straightforward, while
providing a basis for extensions to more complex settings.

\section{Background}
\label{sec:existing}

\subsection{Mixed-effects models encode structure but limit flexibility}

Linear mixed-effects models are a standard tool for longitudinal and
clustered data \citep{LairdWare1982, PinheiroBates2000,
  VerbekeMolenberghs2000, Fitzmaurice2011}. Linear mixed-effects
models may be written in the form
\[
  \mu_i(x) = f_0(x) + \sum_{j=1}^K z_j(x)u_{ij},
\]
where $f_0$ is an unknown shared population trajectory, $z_j(x)$ are
known random-effects design functions (e.g. $z_j(x) = 1$ for a random
intercept, $z_j(x) = x$ for a random slope), and $u_i = (u_{i1},
\ldots, u_{iK})^\T \sim N(0, \Sigma_u)$ are random effects.  Various
models are available for $f_0$: classical linear mixed-effects models
might assume a simple parametric form such as $f_0(x) = \beta_0 +
\beta_1 x$, while generalized additive mixed models (GAMMs)
\citep{Lin1999} allow $f_0$ to be any smooth function of covariates.
GAMMs avoid restrictive assumptions on the shape of the population
mean, but still enforce fixed relationships between individual
trajectories and the mean.  Non-linear mixed-effects models
\citep{LindstromBates1990}
have also
been developed for specific applications \citep[e.g.][for growth
  curves]{Cole2010}, but do not provide general-purpose flexible
models for the trajectories.

Mixed-effects models encode structural similarities between clusters
through a small number of random effects, and therefore meet
requirement~\ref{req:sharing}. AdaStruMMs build on this strength. The
key difference is that standard mixed-effects models require the analyst
to specify the random-effects design functions in advance. This is
natural when the structure is scientifically meaningful, but restrictive
when there is no clear substantive reason to prefer, for example, a
random intercept, random slope, or any other fixed low-dimensional
form. In such cases, the choice of random-effects structure is often
driven by modelling convenience or by the structures available in
standard software. This limits individual-trajectory flexibility, so
these models do not meet requirement~\ref{req:flexible}. AdaStruMMs
preserve the mixed-effects mechanism for sharing information between
clusters, while replacing the pre-specified random-effects design by
functions estimated from the data.

\subsection{Local and hybrid models do not encode structural similarities}

Local approaches model each trajectory $\mu_i$ separately, and
estimate $\mu_i$ from the local data $x_i = (x_{i1}, \ldots
x_{in_i})^\T$ and $y_i = (y_{i1}, \ldots, y_{i n_i})^\T$ alone.  Any
choice of flexible regression model could be used, for instance
smoothing splines or Gaussian processes. If there are a large number
$n_i$ of observations on individual $i$, such local approaches will
provide an accurate estimate of $\mu_i$. However, local approaches
are not suited to clusters with small $n_i$, and local estimates of
$\mu_i(x)$ will be highly uncertain for values of $x$ far from the
local data $x_i$.

Hybrid approaches are similar to local approaches but with some
limited information sharing between clusters.  For instance,
\citet{Pedersen2019} describe various Hierarchical Generalized
Additive Models (HGAMs), including the HGAM-GS, where individual
trajectories are allowed to be arbitrary smooth curves, shrunk towards
a common population trajectory and assumed to be similarly smooth to
one another. Similarly, Gaussian processes could be fitted to each
individual trajectory, with common smoothing parameters and mean
function. These approaches are very flexible, allowing essentially any
smooth trajectory shape (requirement \ref{req:flexible}), but do not
encode structural similarities in clusters beyond their overall level
of smoothness (requirement \ref{req:sharing}).

These approaches with limited information sharing may be close to
optimal when there is little common structure between individuals
beyond smoothness. In many applications, however, there are more
specific structural similarities in the shapes of trajectories, and it
is important to exploit these similarities. This motivates our
development of AdaStruMMs, which aim to identify and exploit these
similarities to improve inference.

\subsection{Sparse FPCA methods can be inaccurate in some cases}
\label{sec:fpca}

Functional principal components analysis (FPCA) was originally used to
study the variation between observed functions. If we observed the
individual trajectories $\mu_i$ directly, we could use FPCA to
identify the dominant modes of variation between trajectories, which
are called functional principal components or eigenfunctions.

This is formalized by the Karhunen--Lo\`eve decomposition. If
$\mu_i$ is a random trajectory with mean function $m$ and
covariance function $C$, then, under standard regularity
conditions,
\begin{equation}
  \label{eq:KL}
  \mu_i(x) = m(x) + \sum_{k=1}^\infty \xi_{ik}\phi_k(x),
\end{equation}
where the $\phi_k$ are orthonormal eigenfunctions of the covariance
operator and the scores $\xi_{ik}$ are uncorrelated random variables
with mean zero and variances $\lambda_k$. FPCA estimates the leading
eigenfunctions $\phi_k$ and the associated scores.

FPCA methods may also be used to estimate the individual trajectories
from clustered data, with two-step ``sparse FPCA'' methods
\citep{Yao2005, Di2009, Xiao2016} or with Bayesian approaches
\citep{Nolan2025}.

FPCA methods are flexible (requirement \ref{req:flexible}) and encode
structural similarities (requirement \ref{req:sharing}), but the
sparse FPCA methods considered here can give inaccurate trajectory
estimates in some cases. In Section \ref{sec:2re_sim}, we give an
example in which sparse FPCA methods sometimes give less accurate
estimates than a model which ignores structural similarities, despite
strong structural similarities in the true trajectories. In this
example, sparse FPCA methods do not exploit these structural
similarities to improve the accuracy of estimates (requirement
\ref{req:works}).

\section{Adaptively-Structured Mixed Models}
\label{sec:model}

\subsection{A mixed-effects model with unknown random-effects design}
\label{sec:adastrumms}

AdaStruMMs are mixed-effects models in which the known random-effects
design functions are replaced by unknown functions estimated from the
data. Under model~\eqref{eq:model_Y}, we model the trajectories as
\begin{align}
\mu_i(x)
  &=
  f_0(x)+\sum_{k=1}^K u_{ik}f_k(x),
  \qquad
  u_{ik}\stackrel{\mathrm{i.i.d.}}{\sim}N(0,1),
  \label{eq:model}\\
f_k(x)
  &=
  \sum_{l=1}^{n_B}\beta_{kl}b_l(x)
  =
  \beta_k^\T b(x),
  \qquad k=0,1,\ldots,K.
  \notag
\end{align}
Here \(b_l(\cdot)\) are fixed cubic spline basis functions,
\(b(x)=(b_1(x),\ldots,b_{n_B}(x))^\T\), and
\(\beta_k=(\beta_{k1},\ldots,\beta_{kn_B})^\T\) is the coefficient
vector for \(f_k\). The unknown quantities are
\(\beta_0,\beta_1,\ldots,\beta_K\) and \(\sigma^2\). For now, we view
\(K\) as fixed, but describe its selection in
Section~\ref{sec:choose_K}. The \(u_{ik}\) and \(\epsilon_{ij}\)
are mutually independent and are independent of the cluster sizes and
covariates.

The assumption that the \(u_{ik}\) are independent standard normal
variables is not restrictive. In a classical mixed-effects model with
random effects \(b_i\sim N_K(0,\Sigma)\), we can write \(b_i=L u_i\),
where \(\Sigma=LL^\T\) and \(u_i\sim N_K(0,I_K)\). The matrix \(L\)
can then be absorbed into the random-effects design functions
\(f_1,\ldots,f_K\).

To make the model identifiable, we impose constraints on
\(\beta_1,\ldots,\beta_K\). Motivated by the Karhunen--Lo\`eve
decomposition~\eqref{eq:KL}, we choose these constraints to make
\(f_1,\ldots,f_K\) orthogonal.

Let \(\mathcal I\) denote the interval over which the spline basis is
constructed, and define
\(\langle g,h\rangle_{\mathcal I}
=\int_{\mathcal I}g(x)h(x)\,dx\).
We use the \texttt{orthogonalsplinebasis} R package
\citep{orthogonalsplinebasis} to construct basis functions satisfying
\(\langle b_i,b_j\rangle_{\mathcal I}=1\) if \(i=j\) and zero
otherwise. It follows that
\(\langle f_i,f_j\rangle_{\mathcal I}=\beta_i^\T\beta_j\).
We therefore impose
\begin{equation}
\label{eq:orthog_constraint}
  \beta_i^\T\beta_j=0,
  \qquad 1\leq i<j\leq K.
\end{equation}
This makes \(f_1,\ldots,f_K\) orthogonal. Because the random effects
have unit variance, the corresponding eigenvalue is
\(\lambda_k=\|f_k\|_{\mathcal I}^2\); equivalently, one could use
orthonormal component functions and random effects with variances
\(\lambda_k\).

When the component norms are non-zero and distinct, the orthogonality
constraint~\eqref{eq:orthog_constraint} makes the model identifiable
up to permutation and sign. We can permute the order of
\(\beta_1,\ldots,\beta_K\), or for any \(k\) replace \(\beta_k\) by
\(-\beta_k\) and each \(u_{ik}\) by \(-u_{ik}\), without altering the
overall process. In principle, we could add constraints to make the
model fully identifiable, for example by ordering the components so
that
\(\|\beta_1\|\geq\|\beta_2\|\geq\cdots\geq\|\beta_K\|\)
and requiring \(\beta_{k1}\geq0\). We do not impose these constraints
during optimization. Instead, after fitting, we permute the estimated
components into decreasing order of norm. Sign changes could similarly
be applied after fitting if needed, but we do not do this in practice.

\subsection{Derive the log-likelihood under the model}

The distribution of \(Y_i=(Y_{i1},\ldots,Y_{in_i})^\T\) depends on
each coefficient vector \(\beta_k\) through \(f_k(x)=\beta_k^\T
b(x)\). For \(k=0,1,\ldots,K\), define \(f_{ki}(\beta_k)
=\{f_k(x_{i1}),\ldots,f_k(x_{in_i})\}^\T =X^{(i)}\beta_k\), where
\(X^{(i)}\) is the \(n_i\times n_B\) basis design matrix for cluster
\(i\).

For optimization, we parameterize the error variance by
\(\eta=\log\sigma\), so that \(\sigma^2=\exp(2\eta)\), and write
\(\beta=(\beta_1,\ldots,\beta_K)\) and
\(\theta=(\beta_0,\beta,\eta)\).

Under model~\eqref{eq:model}, the marginal distribution of \(Y_i\)
is
\begin{align}
Y_i
  &\sim
  N_{n_i}\{m_i(\theta),\Sigma_i(\theta)\},
  \notag\\
m_i(\theta)
  &=f_{0i}(\beta_0),
  \notag\\
\Sigma_i(\theta)
  &=
  \exp(2\eta)I_{n_i}
  +
  \sum_{k=1}^K
  f_{ki}(\beta_k)f_{ki}(\beta_k)^\T.
  \label{eq:Sigma}
\end{align}
The log-likelihood is therefore
\[
  \ell(\theta)
  =
  \sum_{i=1}^d
  \log\phi_{n_i}
  \{y_i;m_i(\theta),\Sigma_i(\theta)\},
\]
where \(\phi_n(\,\cdot\,;\mu,\Sigma)\) is the density of the
\(N_n(\mu,\Sigma)\) distribution.

\subsection{Impose a smoothing penalty to avoid over-fitting}

To avoid over-fitting, we penalize the expected wiggliness of an
individual trajectory. Write
\(w(f)=\int_{-\infty}^{\infty}\{f^{\prime\prime}(x)\}^2\,dx\)
and \(w_E=E\{w(\mu_i)\}\). Penalties based on integrated squared
second derivatives are standard in spline smoothing and
semiparametric regression \citep[e.g.][]{RuppertWandCarroll2003}.

Because the random effects are independent, with mean zero and unit
variance, the expected wiggliness is
\[
  w_E
  =
  w(f_0)+\sum_{j=1}^K w(f_j).
\]
If \(S\) is the usual spline penalty matrix, so that
\(w(f_j)=\beta_j^\T S\beta_j\), this becomes
\(w_E=\sum_{j=0}^K\beta_j^\T S\beta_j\).

The penalized log-likelihood is
\[
  \ell_p(\theta;\gamma,K)
  =
  \ell(\theta)
  -
  \frac{\gamma}{2\sigma^2}
  w_E(\beta_0,\beta_1,\ldots,\beta_K),
\]
where \(\gamma\) is a smoothing parameter. We initially treat
\(\gamma\) as fixed, returning to its selection in
Section~\ref{sec:choose_gamma}, where we also justify scaling the
penalty by \(2\sigma^2\).

\subsection{Define a penalized maximum likelihood estimate}

For fixed \(K\) and \(\gamma\), we maximize the penalized
log-likelihood subject to the orthogonality constraint. Write
\begin{equation}
\label{eq:B_perp}
\begin{aligned}
\mathcal B_\perp
  &=
  \bigl\{
    \beta=(\beta_1,\ldots,\beta_K):
    \beta_j^\T\beta_k=0,\ 1\leq j<k\leq K
  \bigr\},\\
\Theta_\perp
  &=
  \bigl\{
    (\beta_0,\beta,\eta):
    \beta_0\in\mathbb R^{n_B},
    \ \beta\in\mathcal B_\perp,
    \ \eta\in\mathbb R
  \bigr\}.
\end{aligned}
\end{equation}
The penalized maximum likelihood estimator satisfies
\[
  \hat\theta
  \in
  \arg\max_{\theta\in\Theta_\perp}
  \ell_p(\theta;\gamma,K).
\]

This remains a constrained optimization problem. In
Section~\ref{sec:transform}, we introduce a parameterization that
enforces orthogonality automatically.

\subsection{Reparameterize to avoid constrained optimization}
\label{sec:transform}

Directly maximizing over \(\Theta_\perp\) would require constrained
optimization. We instead use an unconstrained working parameterization
that automatically enforces orthogonality.

Write \(\psi=(\beta_0,\alpha,\eta)\), where
\(\alpha=(\alpha_1,\ldots,\alpha_K)\) and
\(\alpha_k\in\mathbb R^{n_B-k+1}\).
Let \(a\) index the possible working parameterizations. For each
\(a\), we construct a map
\[\alpha\mapsto\beta_a(\alpha)\in\mathcal B_\perp,\]
and define
\[
\begin{aligned}
T_a(\psi)
  &=
  \{\beta_0,\beta_a(\alpha),\eta\}
  \in\Theta_\perp,\\
\hat\psi_a
  &\in
  \arg\max_{\psi\in\mathbb R^p}
  \ell_p\{T_a(\psi);\gamma,K\},
  \qquad
  \hat\theta=T_a(\hat\psi_a),
\end{aligned}
\]
where
\(p=n_B+\sum_{k=1}^K(n_B-k+1)+1\).
Thus the orthogonality constraint is enforced automatically while the
optimization is unconstrained.

There are several possible choices of
\(\alpha\mapsto\beta_a(\alpha)\), described in
\suppsec{sec:construct_param}. Each gives a local parameterization of
the same constrained component space, except on a singular set that
depends on \(a\). A value close to a singularity under one
parameterization may be well behaved under another. These
singularities are artefacts of the working parameterization, not the
model. In practice, we therefore allow several working
parameterizations and use one that is well behaved near the fitted
value, as described in \suppsec{sec:choose_param}.

\subsection{Refine the optimization procedure for stability and speed}

We do the fitting sequentially, increasing $K$ one at a time: first
fitting the model with $K = 0$, then increasing $K$ one at a time,
using the parameter values from the previous fit to give starting
values for the optimization.

To increase the speed of optimization, it is important to have access
to the gradient of the penalized log-likelihood, which would be
difficult to find by hand because of the complexity of the
transformation in Section \ref{sec:transform}.  We use automatic
differentiation to obtain the gradient, by using the Stan Math
C\texttt{++} library in R \citep{StanHeaders}.

\subsection{Estimate the individual trajectories}

Given \(\hat\theta\), we estimate the individual trajectory and random
effects by
\[
\begin{aligned}
\hat\mu_i(x)
  &=
  \hat f_0(x)+
  \sum_{k=1}^K\hat u_{ik}\hat f_k(x),
  \qquad
  \hat f_k(x)=\hat\beta_k^\T b(x),\\
\hat u_i
  &=
  \hat F_i^\T\hat\Sigma_i^{-1}
  (y_i-\hat f_{0i}).
\end{aligned}
\]
Here \(\hat F_i\) is the \(n_i\times K\) matrix with columns \(\hat
f_{1i},\ldots,\hat f_{Ki}\), and \(\hat\Sigma_i\) is the marginal
covariance matrix in~\eqref{eq:Sigma}, evaluated at the fitted
parameter. The estimate \(\hat u_i\) is the usual empirical Bayes, or
BLUP, estimate of the random effects under the fitted mixed-effects
model.

\subsection{Compute confidence intervals}
\label{sec:ci}

Each individual trajectory may be written as
\(\mu_i(x)=\delta_i^\T b(x)\), where
\(\delta_i=\beta_0+\sum_{k=1}^K u_{ik}\beta_k\).
We use a local normal approximation in the working parameterization to
generate a sample
\(\delta_i^{(1)},\ldots,\delta_i^{(n_S)}\), as described in
\suppsec{sec:ci_details}. A pointwise 95\% confidence interval for
\(\mu_i(x)\) is obtained from the 2.5th and 97.5th percentiles of
\[
  \bigl\{
    \{\delta_i^{(j)}\}^{\T}b(x):
    j=1,\ldots,n_S
  \bigr\}.
\]
The same sample can be used for derived quantities; for example,
replacing \(b(x)\) by \(b^\prime(x)\) gives intervals for
\(\mu_i^\prime(x)\).

This procedure treats \(K\) and \(\gamma\) as fixed and may therefore
lead to undercoverage. Nevertheless, coverage is close to nominal in
the simulation studies.

\subsection{Choose the number of components \(K\)}
\label{sec:choose_K}

For fixed \(\gamma\), the fit eventually stabilizes as \(K\)
increases, with \(\hat\lambda_K=\|\hat f_K\|^2\) approaching zero. We
therefore choose \(K\) as small as possible while retaining the
important components.

For each candidate \(K\), let \(\hat\sigma_K^2\) denote the estimated
error variance. We choose the smallest \(K\) for which
\[
  \operatorname{FVE}(K;K_{\max})
  =
  \frac{\hat\sigma_0^2-\hat\sigma_K^2}
       {\hat\sigma_0^2-\hat\sigma_{K_{\max}}^2}
  >
  t_{\mathrm{FVE}},
\]
where \(K_{\max}\) is increased adaptively until
\(\operatorname{FVE}(K_{\max}-1;K_{\max})>t_{\mathrm{FVE}}\).
Full details of the iterative procedure are given in
\suppsec{sec:choose_K_details}. We use
\(t_{\mathrm{FVE}}=0.999\) throughout.

\subsection{Choose the smoothing parameter \(\gamma\)}
\label{sec:choose_gamma}

We choose \(\gamma\) to maximize an approximate marginal likelihood
based on the usual Bayesian interpretation of smoothing penalties
\citep[Section~6.2.6]{Wood2017}. For fixed \(\gamma\) and \(K\), with
\(\sigma\) fixed at its penalized maximum likelihood estimate, the
criterion is
\begin{equation}
\begin{split}
  \log\tilde m_{\gamma,K}
  ={}&
  \ell_p(\hat\theta_{\gamma,K})
  +\frac12\log|P|_{+,\hat G}
  -\frac12\log|H|_{\hat G}\\
  &+\frac M2\log(2\pi)-\log(K!2^K).
\end{split}
\label{eq:approx_laml_gamma}
\end{equation}
Here \(P\) is the local penalty precision, \(H\) is the negative
penalized-likelihood Hessian, and \(\hat G\) is the metric induced by
the working parameterization. The quantity \(M\) is the dimension of
the unpenalized subspace, equivalently the dimension of the null space
of \(P\). The symbols \(|\cdot|_{\hat G}\) and
\(|\cdot|_{+,\hat G}\) denote generalized determinants relative to
\(\hat G\), with the latter including only positive generalized
eigenvalues. All terms in the criterion are fully defined, and its
derivation is given in \suppsec{sec:gamma_choice_details}.

For computational reasons, we first choose \(K\) separately for each
candidate \(\gamma\), then evaluate the criterion once for that value.
This avoids repeatedly computing the Hessian during the search over
\(K\).

\subsection{Estimate the population-level process}
\label{sec:pop_inference}

Averaging model~\eqref{eq:model} over \(u_i\) gives a population
Gaussian process with
\begin{align}
m(x)
  &=E\{\mu_i(x)\}=f_0(x),
  \notag\\
C(x,x')
  &=
  \operatorname{cov}\{\mu_i(x),\mu_i(x')\}
  =
  \sum_{k=1}^K f_k(x)f_k(x').
  \label{eq:cov_fpc}
\end{align}
The corresponding plug-in estimators are
\[
  \hat m(x)=\hat f_0(x),
  \qquad
  \hat C(x,x')
  =
  \sum_{k=1}^K\hat f_k(x)\hat f_k(x').
\]

For comparison, we also construct empirical mean and covariance
estimators from the fitted individual trajectories:
\[
\begin{aligned}
\tilde m(x)
  &=
  d^{-1}\sum_{i=1}^d\hat\mu_i(x),\\
\tilde C(x,x')
  &=
  d^{-1}\sum_{i=1}^d
  \{\hat\mu_i(x)-\tilde m(x)\}
  \{\hat\mu_i(x')-\tilde m(x')\}.
\end{aligned}
\]
These estimators do not account for the shrinkage of the fitted
trajectories towards the population mean, so the empirical covariance
estimator may underestimate between-individual variation.

We therefore also consider a corrected estimator. Write
\(\hat{\boldsymbol f}(x)
=(\hat f_1(x),\ldots,\hat f_K(x))^\T\), and let \(\hat V_i\) be the
estimated conditional covariance matrix of \(u_i\). The
empirical-corrected estimator uses the same mean \(\tilde m\), but
replaces \(\tilde C\) by
\[
  \tilde C_{\mathrm{corr}}(x,x')
  =
  \tilde C(x,x')
  +
  \frac{1}{d}\sum_{i=1}^d
  \hat{\boldsymbol f}(x)^\T
  \hat V_i
  \hat{\boldsymbol f}(x').
\]
The additional term restores the average conditional variation omitted
when the covariance is calculated using only the fitted conditional
mean trajectories.

For AdaStruMM population-level inference in the simulation studies, we
use the empirical-corrected estimators, which gave slightly more
accurate estimates than the plug-in estimators in preliminary
comparisons. Further details of this choice are given in
\suppsec{sec:simstudies_gp_choice}. The population-level asymptotic results
in Theorem~\ref{thm:pop} concern the plug-in estimators.

\section{Asymptotic theory}
\label{sec:theory}

\subsection{Introduction}

As the number of clusters \(d\) tends to infinity, with a bounded
number of observations per cluster, we establish asymptotic properties
of AdaStruMMs at both population and individual level.

At population level, suppose that model~\eqref{eq:model} is correctly
specified with \(K=K_0\) components, and that AdaStruMM is fitted with
\(K=K_0\) and \(\gamma\) fixed at any value. Standard likelihood
arguments then apply. Theorem~\ref{thm:MLE} establishes
consistency of the canonical fitted parameter and asymptotic normality
in a regular local working parameterization. As a consequence,
Theorem~\ref{thm:pop} establishes consistency and asymptotic normality
for the induced estimators of the population mean and covariance
functions.

At individual level, Theorem~\ref{thm:traj} shows that AdaStruMM
estimators of individual trajectories are asymptotically equivalent to
oracle estimators obtained with the model parameter fixed at its true
value. Thus, estimating the random-effects design incurs
asymptotically no additional individual-level prediction error.

\subsection{Define a canonical parameter for theoretical developments}
\label{sec:canonical_parameter}

As discussed in Section~\ref{sec:adastrumms}, the model is
identifiable only up to permutation and sign changes of the component
coefficient vectors \(\beta = (\beta_1,\ldots,\beta_K)\). For
theoretical developments, we therefore work with a canonical
representative of each equivalence class.

Given \(\theta=(\beta_0,\beta,\eta)\in\Theta_\perp\), define its
canonical version \(\theta^*=(\beta_0,\beta^*,\eta)\) as follows. First
choose the sign of each component by defining
\[
  \tilde\beta_k =
  \begin{cases}
    -\beta_k, & \beta_{k1}<0,\\
    \beta_k, & \beta_{k1}\geq 0.
  \end{cases}
\]
Next, let \(\pi\) be a permutation such that
\[
  \|\tilde\beta_{\pi(1)}\|
  \geq
  \|\tilde\beta_{\pi(2)}\|
  \geq
  \cdots
  \geq
  \|\tilde\beta_{\pi(K)}\|,
\]
where \(\|\cdot\|\) denotes the Euclidean norm. We then define
\[
  \beta^*
  =
  (\tilde\beta_{\pi(1)},\ldots,\tilde\beta_{\pi(K)}).
\]
Let \(\Theta_\perp^*\) denote the set of canonical representatives
obtained in this way.

The canonical representative is not unique on boundary sets.  If
\(\beta_{k1}=0\) for some \(k\), the sign convention does not
distinguish \(\beta_k\) from \(-\beta_k\); if
\(\lVert\beta_j\rVert=\lVert\beta_k\rVert\) for some \(j\ne k\), the
ordering convention does not distinguish the two components. The
quantities of scientific interest considered below, such as the
population mean and covariance functions, are invariant to these sign
and ordering choices. However, to obtain a simple asymptotic normality
statement for the parameter estimator itself, we assume below that the
true canonical parameter is separated from these boundary sets.

\subsection{Use a local working parameterization for asymptotic normality}

We state consistency directly for a canonical model parameter in
\(\Theta_\perp^*\). To express its asymptotic distribution in ordinary
Euclidean coordinates, we use one of the working parameterizations from
Section~\ref{sec:transform} only locally around the true parameter.

Let \(\theta_0\) denote the true parameter and let \(\theta_0^*\) denote
its canonical version. Fix \(a\in\mathcal A\), and let \(\psi_0^*\) be
the corresponding working parameter, so that
\[
  \theta_0^* = T_a(\psi_0^*).
\]
Assumption~\ref{ass:non_singular} below
requires this parameterization to be non-singular at \(\psi_0^*\).
Under that assumption, \(T_a\) gives a smooth, one-to-one local
parameterization around \(\theta_0^*\), with a smooth local inverse.
The practical fitting procedure need not use this fixed
parameterization throughout: it may switch parameterizations during
optimization and canonicalize the fitted components afterwards.

\subsection{Impose regularity conditions}

We impose the following regularity conditions:
\begin{enumerate}[label=(A\arabic*),ref=A\arabic*]

\item \emph{Bounded random cluster sizes.}
\label{ass:n_i}
The cluster sizes \(n_1,\dots,n_d\) are independent and identically
distributed integer-valued random variables satisfying
\(1 \leq n_i \leq n_{\max} < \infty\) almost surely and
\(\operatorname{pr}(n_i \geq 2)>0\).

\item \emph{Random design with interval support.}
\label{ass:x}
Conditional on \(n_i\), the covariates
\(x_{i1},\ldots,x_{in_i}\) are independent and identically distributed
from \(P_X\), independently across clusters. The distribution \(P_X\) has support
\[
  \mathcal I=[a,b], \qquad a<b,
\]
where \(\mathcal I\) is the interval over which the spline basis is
constructed.

\item \emph{Compact parameter space.}
  \label{ass:not_boundary}
For fixed \(K=K_0\), let
\(\Theta_c^*\subset\Theta_\perp^*\) be a fixed compact parameter
space. Assume that there exists \(\varepsilon>0\) such that
\[
  \left\{
    \theta\in\Theta_\perp^*:
    \|\theta-\theta_0^*\|<\varepsilon
  \right\}
  \subseteq \Theta_c^*.
\]
Assume that every
\(\theta=(\beta_0,\beta_1,\ldots,\beta_{K_0},\eta)\in\Theta_c^*\)
satisfies
\[
  \beta_{k1}>0,\qquad k=1,\ldots,K_0,
\]
and
\[
  \|\beta_1\|>\|\beta_2\|>\cdots>\|\beta_{K_0}\|>0.
\]

\item \emph{True parameter not in the singular set of the local
working parameterization.}
\label{ass:non_singular}
Assume that
\[
  \psi_0^*\notin S_a,
\]
where \(S_a\) is the singular set for the chosen working
parameterization, defined in \eqref{eq:singular_set}.

\end{enumerate}

Assumption~\ref{ass:non_singular} is needed only to express the local
asymptotic distribution; if it fails for one choice of \(a\), a
different working parameterization may be
used. Lemma~\ref{lemma:information_nonsingular} in
\suppsec{app:proofs} shows that the corresponding Fisher
information is nonsingular under
Assumptions~\ref{ass:n_i}--\ref{ass:non_singular}.

\subsection{Asymptotic results at population level}

The first result establishes consistency of the fitted canonical
parameter and asymptotic normality in a regular local working
parameterization.

\begin{theorem}
\label{thm:MLE}
Suppose that Assumptions~\ref{ass:n_i}--\ref{ass:not_boundary} hold,
and that the model is correctly specified with \(K=K_0\). Let
\(\hat\theta^*\) be a maximizer of the penalized log-likelihood over
the fixed compact canonical parameter space, so that
\[
  \hat\theta^*
  \in
  \arg\max_{\theta\in\Theta_c^*}
  \ell_p(\theta;\gamma,K_0),
\]
where \(\gamma\) is fixed. Then, as \(d\to\infty\),
\[
  \hat\theta^* \xrightarrow{p} \theta_0^*.
\]

If, in addition, Assumption~\ref{ass:non_singular} holds for a fixed
working parameterization \(a\), consistency implies that, with
probability tending to one, \(\hat\theta^*\) lies in the image of the
local parameterization around \(\theta_0^*\). On this event define
\[
  \hat\psi^* = T_a^{-1}(\hat\theta^*).
\]
Then
\[
  d^{1/2}(\hat\psi^*-\psi_0^*)
  \xrightarrow{d}
  N\!\left\{0, I_a(\psi_0^*)^{-1}\right\},
\]
where \(I_a(\psi)\) is the Fisher information matrix for a single
cluster in the fixed local working parameterization,
\[
  I_a(\psi)
  =
  E_\psi
  \left[
    -\nabla_\psi^2
    \ell_i\{T_a(\psi);Y_i\mid n_i,x_i\}
  \right],
\]
with expectation taken over the joint distribution of
\((n_i,x_i,Y_i)\).
\end{theorem}
All proofs are given in \suppsec{app:proofs}.

The next result transfers these properties to the plug-in estimators
of the population mean and covariance functions.

\begin{theorem}
\label{thm:pop}
Suppose that Assumptions~\ref{ass:n_i}--\ref{ass:not_boundary} hold,
that the model is correctly specified with \(K=K_0\), and that
\(\hat\theta^*\) is the penalized maximum likelihood estimator defined
in Theorem~\ref{thm:MLE}, with \(\gamma\) fixed. Let
\[
  m(x) = f_0(x), \qquad
  C(x,x') = \sum_{k=1}^{K_0} f_k(x)f_k(x'),
\]
denote the population mean and covariance functions, and let
\[
  \hat m(x) = \hat f_0(x), \qquad
  \hat C(x,x') =
  \sum_{k=1}^{K_0} \hat f_k(x)\hat f_k(x'),
\]
be the corresponding plug-in estimators. Then, for each fixed
\(x,x'\in\mathcal I\),
\[
  \hat m(x) \xrightarrow{p} m(x),
  \qquad
  \hat C(x,x') \xrightarrow{p} C(x,x').
\]

If, in addition, Assumption~\ref{ass:non_singular} holds for a fixed
working parameterization \(a\), then
\[
  d^{1/2}\{\hat m(x)-m(x)\}
  \xrightarrow{d}
  N\{0,\tau_m^2(x)\},
\]
and
\[
  d^{1/2}\{\hat C(x,x')-C(x,x')\}
  \xrightarrow{d}
  N\{0,\tau_C^2(x,x')\}.
\]
The asymptotic variances \(\tau_m^2(x)\) and
\(\tau_C^2(x,x')\) are the delta-method variances defined in
\eqref{eq:tau_m_def} and \eqref{eq:tau_C_def} in
\suppsec{app:proofs}.
\end{theorem}

\subsection{Asymptotic results at individual level}

The final result shows that estimating the population-level structure
has an asymptotically negligible effect on prediction of any fixed
individual trajectory.

\begin{theorem}\label{thm:traj}
Suppose that Assumptions~\ref{ass:n_i}--\ref{ass:not_boundary} hold,
and that the model is correctly specified with \(K=K_0\). Let
\(\hat\theta^*\) be the estimator defined in
Theorem~\ref{thm:MLE}. For cluster \(i\), let
\[
  \hat\mu_i(x)
  =
  E_{\hat\theta^*}\{\mu_i(x)\mid y_i,x_i\},
\]
denote the AdaStruMM estimator of the trajectory, and let
\[
  \mu_i^{\mathrm{or}}(x)
  =
  E_{\theta_0^*}\{\mu_i(x)\mid y_i,x_i\},
\]
denote the corresponding oracle predictor under the true parameter.
Then, for each fixed \(i\) and \(x\in\mathcal I\),
\[
  \hat\mu_i(x)-\mu_i^{\mathrm{or}}(x)
  \xrightarrow{p}0.
\]
\end{theorem}

\section{Simulation studies}
\label{sec:sim_studies}

\subsection{Design of the simulation studies}

\medskip
\noindent\textit{Data-generating processes.}\label{sec:sim_generation}
To compare methods empirically, we simulate from model
\eqref{eq:model_Y}, with a range of choices for the true trajectories
\(\mu_i\). The main study uses two data-generating processes:
\begin{enumerate}
\item \emph{Two-component model.} A mixed-effects model with
  $K = 2$ components.  We let $\mu_i(x) = h(x) + u_{i1} h(x) +
  u_{i2}$, where $h(x) = x/2 + \sin(x)$ and $u_{i1}$ and $u_{i2}$ are
  independent $N(0, 1)$ random variables. We generate $x_{ij}$
  uniformly on the interval $[0, 3\pi]$. The measurement error 
  standard deviation is $\sigma = 0.1$. Results are given in Section
  \ref{sec:2re_sim}.
\item \emph{Growth curve model.} A non-linear random effects model
  for growth curves called SuperImposition by Translation And Rotation
  (SITAR) \citep{Cole2010}, in which
\[
  \mu_i(x) = \alpha_i + h\left(\frac{x -
    \beta_i}{\exp(-\gamma_i)}\right),
\]
where $\alpha_i, \beta_i, \gamma_i$ are random effects, and $h$ is
an unknown function which must be estimated. 

In order to simulate data from a realistic process, we first fit a
SITAR model to heights of girls aged 8--18, from the Berkeley Child
Guidance Study dataset \citep{Tuddenham1954}, then simulate from the
fitted model. The data are available as \texttt{berkeley} in the
\texttt{sitar} R package \citep{sitar}, and the subset we use is also
used in the package vignette. For each individual, the first observation
point $x_{i1}$ is uniformly distributed on $[8, 8+h]$, where $h =
(18-8)/n_i$, with the remaining observations $x_{ij} = x_{i1} +
(j-1)h$ spaced $h$ years apart.

We generate data from SITAR as an example of a realistic process which
is not exactly expressible in the form of the model \eqref{eq:model}
for any $K$, but for which there are nonetheless strong structural
similarities between trajectories. In the example we consider, 99.7\%
of the variability in trajectories is explained by $K = 3$ components.
Results are given in Section \ref{sec:sitar_sim}.
\end{enumerate}

We study the impact of varying the number of individuals, $d$, and the
number of observations on each individual, $n_i$. For each case, we
generate 100 datasets from each of 24 processes, given by all possible
combinations of $d \in \{50, 100, 200, 300, 400, 500\}$ and
$n_i \in \{2, 3, 5, 10\}$ (with an equal number of observations on each
individual).

We also consider an additional correctly specified model with three
random-effects components. Its specification and results are reported
in \suppsec{sec:three_component_simulation}; the conclusions are
qualitatively similar to those for the two-component case.

\medskip
\noindent\textit{Compare six general-purpose methods.}\label{sec:sim_methods}
We compare six general-purpose estimation methods:
\begin{enumerate}
\item \emph{AdaStruMM.} The proposed method, with \(K\) and
  \(\gamma\) chosen as described in Sections~\ref{sec:choose_K} and
  \ref{sec:choose_gamma}, using \(t_{\mathrm{FVE}}=0.999\).
\item \emph{Local-GAM.} A separate smooth spline fitted to each
  cluster, included as a simple local baseline and reported only for
  \(n_i\geq5\).
\item \emph{HGAM-GS.} A hierarchical generalized additive model with
  a common population mean and shared smoothing parameters
  \citep{Pedersen2019}.
\item \emph{PACE.} The sparse functional principal components method
  of \citet{Yao2005}, retaining components that explain 99\% of the
  estimated variation and using generalized cross-validation for
  smoothing.
\item \emph{FACE.} The sparse functional principal components method
  of \citet{Xiao2016}, retaining components that explain 99\% of the
  estimated variation.
\item \emph{bayesFPCA.} The variational message passing method of
  \citet{Nolan2025}, supplied with the true \(K\) as a favourable
  benchmark in the correctly specified cases and omitted from the
  growth-curve case.
\end{enumerate}

In the growth-curve case, we additionally fit the correctly specified
SITAR model \citep{Cole2010,sitar}. This is an application-specific
comparator rather than a general-purpose method.

For population-level inference, we use the empirical-corrected
estimators for AdaStruMM, the plug-in estimators for PACE, FACE and
bayesFPCA, and the empirical estimators for Local-GAM, HGAM-GS and
SITAR. These choices are motivated in
\suppsec{sec:simstudies_gp_choice}.

The preliminary comparison used to fix tuning choices and
population-level estimators, together with the methods not taken
forward and the computational feasibility criteria, is described in
\suppsec{sec:simstudies_pilot}. Exact software implementations and
remaining tuning choices are given in
\suppsec{sec:simstudies_implementation}.

\medskip
\noindent\textit{Construct oracle bounds for comparison.}\label{sec:oracle_bounds}
Finite data impose unavoidable error at both individual and population
level. Even if the population process were known, individual
trajectories would remain uncertain because each is observed only
\(n_i\) times. Even if the true individual trajectories were observed,
population summaries would remain subject to finite-\(d\) variation.
We define two oracle bounds to quantify these sources of error.

For the individual-level oracle bound, we estimate each trajectory
under the true population model, conditional on that individual's
observed data. For the two-component case and the supplementary
three-component case, this uses the known mixed-effects structure; for
the growth-curve case, it uses the known SITAR population model. This
quantifies the unavoidable individual-level error when the population
structure is known. For interval comparisons, the corresponding
individual-level oracle interval is the pointwise conditional 95\%
interval for \(\mu_i(x)\) under the true population model.

For the population-level oracle bound, we calculate the empirical mean
and covariance of the \(d\) true simulated trajectories. This
quantifies the finite-\(d\) error that remains even if every trajectory
were observed without estimation error.

\medskip
\noindent\textit{Compare individual- and population-level inference.}\label{sec:sim_comparisons}
For an estimated trajectory \(\hat\mu_i\), define
\[
  \operatorname{ISE}(\hat\mu_i)
  =
  \int_{x_{\min}}^{x_{\max}}
  \{\hat\mu_i(x)-\mu_i(x)\}^2\,dx,
\]
where \(x_{\min}=\min_{i,j}x_{ij}\) and
\(x_{\max}=\max_{i,j}x_{ij}\). For each simulated dataset, we average
this quantity over individuals to obtain the mean integrated squared
error. For each setting and method, we report the root mean integrated
squared error, defined as the square root of the mean of these values
over the 100 simulated datasets.

We also report average pointwise coverage and width of the 95\%
confidence intervals for the individual trajectories over
\([x_{\min},x_{\max}]\), averaged across the 100 datasets. Details of
how occasional fitting or summary failures are handled, and of the
omitted Local-GAM interval summaries, are given in
\suppsec{sec:sim_evaluation_details}.

Population-level error is measured using the squared 2-Wasserstein
distance between the estimated Gaussian process and a Gaussian
reference process with the true mean and covariance functions, as
defined in \suppsec{sec:sim_comparisons_pop}. In the correctly
specified cases, this reference process is the true population
process. We report the square root of the mean of the squared distance
across the 100 simulation runs.

\subsection{Results of the simulation studies}

\medskip
\noindent\textit{AdaStruMM gives the best inference for the two-component model.}\label{sec:2re_sim}

\begin{figure}
  \begin{subfigure}{\textwidth}
    \centering
    \resizebox{0.96\linewidth}{!}{\input{figures/RMISE-2re.tex}}
    \caption{Root mean integrated squared error
      in individual trajectories.}\label{fig:RMISE-2re}
  \end{subfigure}
 \begin{subfigure}{\textwidth}
    \centering
    \resizebox{0.96\linewidth}{!}{\input{figures/coverage-2re.tex}}
    \caption{Coverage of confidence
      intervals for individual trajectories (nominal level 0.95).}\label{fig:coverage-2re}
 \end{subfigure}
     \begin{subfigure}{\textwidth}
    \centering
    \resizebox{0.96\linewidth}{!}{\input{figures/ci-width-2re.tex}}
    \caption{Average width of confidence
      intervals for individual trajectories.}\label{fig:ci-width-2re}
     \end{subfigure}
     \begin{subfigure}{\textwidth}
    \centering
    \resizebox{0.96\linewidth}{!}{\input{figures/rmw2-2re.tex}}
    \caption{Root mean squared 2-Wasserstein
      error in population-level process.}\label{fig:rmw2-2re}
     \end{subfigure}
  \caption{Summary of two-component
    simulations, comparing AdaStruMM (\legAdaStruMM), Local-GAM
    (\legLocalGAM), HGAM-GS (\legHGAM), PACE (\legPACE), FACE
    (\legFACE) and bayesFPCA (\legBayesFPCA). Dashed lines show oracle
    benchmarks; in panel (b), the dotted line marks nominal 0.95
    coverage. Local-GAM is shown for $n_i \geq 5$, and its
    confidence-interval summaries are omitted. Points and shaded bands
    are estimates and 95\% confidence intervals for each quantity,
    based on 100 simulation runs. The plots are split by number of
    observations per individual, $n_i$. The $x$-axis is the number of
    individuals, $d$.  In some panels, the $y$-axis is truncated for
    readability. }\label{fig:2re-comparison}
\alttext{Panel (a): AdaStruMM has the lowest non-oracle individual-trajectory error and approaches the oracle as the sample sizes increase. Panel (b): coverage is close to 0.95. Panel (c): its intervals are the narrowest among the non-oracle methods. Panel (d): it has the lowest non-oracle population-level error.}
\end{figure}

For the two-component model, AdaStruMM gives the strongest overall
performance across the criteria in Fig.~\ref{fig:2re-comparison}. It
has the lowest individual-trajectory errors, the lowest
population-level errors, close-to-nominal interval coverage, and
substantially narrower intervals than the other non-oracle methods.
AdaStruMM is also substantially faster than the methods closest to it
in statistical performance. In the largest two-component setting,
with \(d=500\) and \(n_i=10\), its average fitting time was
\(0.2\) minutes, compared with \(24\) minutes for bayesFPCA and
\(40\) minutes for HGAM-GS. Full timing results are reported in
\suppsec{sec:simulation_timing}.

AdaStruMM is the only method considered here which meets all our
modelling requirements: it is flexible, learns shared structure, and
exploits this to improve the accuracy of inference. FPCA-based methods
are flexible and estimate shared structure, but this does not always
lead to more accurate inference. For \(n_i = 10\), FPCA methods are
sometimes outperformed by the simple Local-GAM baseline, which fits a
separate model to each cluster.

AdaStruMM performs well even in the sparsest settings. With only two
or three observations per individual, model fitting can be challenging
even for classical mixed-effects models, with the random-effects
structure fixed in advance. AdaStruMM is solving a harder problem
still, since the random-effects structure is unknown and must be
learned from the data. Its accuracy in these settings is therefore a
particularly strong indication that the shared structure is being
learned effectively.

The empirical behaviour of AdaStruMM closely matches the behaviour
suggested by the asymptotic theory. At individual level, in this
correctly specified setting, Theorem~\ref{thm:traj} implies that, for
fixed \(n_i\), the additional error from estimating the
population-level structure should vanish as \(d\) increases: the
AdaStruMM trajectory predictor should approach the oracle predictor
computed with the true population parameter. This is what we see in
Fig.~\ref{fig:2re-comparison}(\subref{fig:RMISE-2re}). As \(d\) increases, the AdaStruMM
individual-level error approaches the individual-level oracle
bound. Once this bound is reached, the remaining error is not due to
estimating the random-effects design, but to the finite number of
observations on each individual trajectory. The interval widths in
Fig.~\ref{fig:2re-comparison}(\subref{fig:ci-width-2re}) show the same pattern empirically. For
\(n_i\geq 5\), AdaStruMM already closely matches the individual-level
oracle even for the smallest \(d\) under study; no other method
considered here shows this behaviour.

At population level, Theorem~\ref{thm:pop} establishes consistency and
asymptotic normality for the plug-in estimator. The
empirical-corrected estimator used in the simulations shows the same
qualitative pattern: for fixed \(n_i\), the AdaStruMM population-level
error decreases as \(d\) increases. The gap between AdaStruMM and the
population-level oracle is small for \(n_i\geq 3\), showing that
little error remains beyond the finite-\(d\) variation captured by the
oracle benchmark; again, no other method considered here shows this
behaviour.

\medskip
\noindent\textit{AdaStruMM gives the best general-purpose inference for the growth curve model.}\label{sec:sitar_sim}

\begin{figure}
  \begin{subfigure}{\textwidth}
    \centering
    \resizebox{0.96\linewidth}{!}{\input{figures/RMISE-sitar.tex}}
    \caption{Root mean integrated squared error
      in individual trajectories.}\label{fig:RMISE-sitar}
  \end{subfigure}
  \begin{subfigure}{\textwidth}
    \centering
    \resizebox{0.96\linewidth}{!}{\input{figures/coverage-sitar.tex}}
    \caption{Coverage of confidence
      intervals for individual trajectories (nominal level 0.95).}\label{fig:coverage-sitar}
  \end{subfigure}
  \begin{subfigure}{\textwidth}
    \centering
    \resizebox{0.96\linewidth}{!}{\input{figures/ci-width-sitar.tex}}
    \caption{Average width of confidence
      intervals for individual trajectories.}\label{fig:ci-width-sitar}
  \end{subfigure}
  \begin{subfigure}{\textwidth}
    \centering
    \resizebox{0.96\linewidth}{!}{\input{figures/rmw2-sitar.tex}}
    \caption{Root mean squared 2-Wasserstein error in population-level process.}\label{fig:rmw2-sitar}
  \end{subfigure}
  \caption{Summary of growth-curve
    simulations, comparing AdaStruMM (\legAdaStruMM), Local-GAM
    (\legLocalGAM), HGAM-GS (\legHGAM), PACE (\legPACE), FACE
    (\legFACE) and SITAR (\legSITAR). Dashed lines show oracle
    benchmarks; in panel (b), the dotted line marks nominal 0.95
    coverage. Local-GAM is shown for $n_i \geq 5$, and its
    confidence-interval summaries are omitted. SITAR is shown for $n_i
    \geq 5$, and SITAR intervals were unavailable. Points and shaded
    bands are estimates and 95\% confidence intervals for each
    quantity, based on 100 simulation runs. The plots are split by
    number of observations per individual, $n_i$. The $x$-axis is the
    number of individuals, $d$. In some panels, the $y$-axis is
    truncated for readability.}\label{fig:sitar-comparison}
\alttext{Panel (a): SITAR performs best where available, while AdaStruMM has the lowest error among the general-purpose methods. Panel (b): AdaStruMM coverage is close to nominal. Panel (c): its intervals are comparatively narrow. Panel (d): it has the lowest population-level error among the general-purpose methods.}
\end{figure}

In the growth-curve simulations, the data are generated from a SITAR
model, so in addition to the general-purpose methods we also fit the
correctly specified SITAR model itself. SITAR is an
application-specific model rather than a general-purpose method, but
it provides an important benchmark for what can be achieved when the
correct structural form is known.  When SITAR fits reliably, it
typically gives the lowest individual-trajectory errors, as
expected. However, SITAR fitting is often unavailable or unreliable in
the sparsest settings ($n_i \leq 3$), where AdaStruMM continues to
provide stable inference.

Among the general-purpose methods, AdaStruMM gives the strongest overall
performance in Fig.~\ref{fig:sitar-comparison}. It gives the most
accurate individual trajectories, close-to-nominal interval coverage,
and the most accurate population-level estimates. AdaStruMM remains
computationally competitive in this misspecified setting; full timing
comparisons are given in \suppsec{sec:simulation_timing}.

In this case AdaStruMM is misspecified and the asymptotic theory is
not directly applicable. Nevertheless, the oracle comparisons show a
similar qualitative pattern to the two-component case and the
supplementary three-component case. At
individual level, larger \(n_i\) reduces AdaStruMM trajectory error,
and increasing \(d\) tends to bring the estimates closer to the oracle.
At population level, the error decreases with \(d\), while the gap to
the oracle is generally smaller for larger \(n_i\). This suggests that
the learned low-dimensional structure captures much of the useful
population information, even without imposing the exact SITAR form.

\section{Application to body fat data}
\label{sec:fat_analysis}

We illustrate the method using body fat measurements from the MIT
Growth and Development Study \citep{Phillips2003}. The dataset is used
in \citet{Fitzmaurice2011} and is available as \texttt{fat} in the
associated ALA R package \citep{ALA}. The package notes that these data
``represent a subset of the study materials and should not be used to
draw substantive conclusions''. Accordingly, we use them only to
demonstrate the methodology and do not interpret the results as
substantive scientific findings.

The data contain measurements of percent body fat for 162 girls, in the
years before and after menarche (time of first period). Measurements
were taken roughly annually, with the last appointment scheduled four
years after menarche. Time is rescaled to be relative to time of
menarche, so that zero represents reported time of menarche. The number
of observations on each individual ranges from 3 to 10, with a mean of
6.5.

We fit an AdaStruMM to these data. The fitted model has \(K=4\)
components and an estimated error standard deviation of
\(\hat\sigma=0.32\).
Figure~\ref{fig:fat_fitted_curves} shows estimated percent body fat
\(\hat\mu_i(x)\) for the first twenty girls, with 95\% confidence
intervals. The fit appears reasonable, but to understand how percent
body fat varies with time in the population as a whole, we need to
summarize beyond the individual fitted curves.

\begin{figure}
\centering \input{figures/fat_fitted_curves.tex}
\caption{Percentage body fat against
  time, for the first twenty girls in the \texttt{fat} data, with
  fitted curves and 95\% confidence intervals overlaid.}\label{fig:fat_fitted_curves}
\alttext{Most individual body-fat trajectories rise around and after menarche, but their levels and shapes vary. Uncertainty is greatest near the ends of the observed time range and for sparsely observed girls.}
\end{figure}

The fitted population mean increases rapidly from six months before
menarche until two years afterwards. Figure~\ref{fig:fat_pa} shows this
mean with confidence intervals, together with the population mean from
the piecewise-linear model of \citet{Fitzmaurice2011}. The simpler
model cannot represent the curvature in the AdaStruMM estimate,
illustrating the benefit of allowing the population mean to be
estimated flexibly.

\begin{figure}
\centering \input{figures/fat_pa.tex}
\caption{The estimated population mean of percent
  body fat over time, with 95\% confidence intervals. The dashed line
  shows the estimated population mean under the piecewise linear model
  of \citet{Fitzmaurice2011}.}\label{fig:fat_pa}
\alttext{Estimated mean body fat dips slightly before menarche, then rises rapidly for about two years afterwards. The smooth estimate differs from the piecewise-linear fit mainly around the dip and in its post-menarche curvature.}
\end{figure}

The population mean does not describe variation between individuals.
The fitted AdaStruMM also quantifies this variation while estimating,
rather than pre-specifying, the ways in which individual trajectories
differ from the mean. Figure~\ref{fig:fat_estimated_derivs} summarizes
the distribution of estimated rates of change, showing the median and
bands covering the central 50\% and 90\% of individuals. The fitted
trajectories generally increase from shortly before menarche until two
years afterwards, with greater between-individual variation before
menarche, illustrating how the method can reveal features of individual
variation without imposing their form in advance.

\begin{figure}
\centering \input{figures/fat_estimated_derivs.tex}
\caption{The distribution of the
  estimated rate of change in percent body fat across individuals. The
  solid line shows the median; the darker and lighter shaded regions
  span the 25th--75th and 5th--95th percentiles, respectively.}\label{fig:fat_estimated_derivs}
\alttext{Median body-fat change is negative before menarche, peaks positively shortly afterwards, then declines while remaining positive. Wide percentile bands show substantial between-individual variation, especially around menarche.}
\end{figure}

\section{Discussion}
\label{sec:discussion}

The results show that learning the random-effects design functions can
retain the familiar mixed-effects mechanisms for information sharing,
shrinkage and individual prediction while avoiding a pre-specified
low-dimensional structure. Extensions to multiple covariates,
non-normal responses and more complex clustered-data structures,
including nested or crossed clustering, are important directions for
future work. These extensions are needed before AdaStruMMs can provide
a general alternative to pre-specified random-effects structures in
applied mixed-effects modelling.

\supplementarymaterialstatement

\section*{Data and code availability}

The body-fat data are available in the \texttt{ALA} R package
\citep{ALA}. The Berkeley growth data used to construct the
growth-curve simulations are available as \texttt{berkeley} in the
\texttt{sitar} R package \citep{sitar}. The \texttt{adastrumm} R
package implementing the
proposed methods, together with code for reproducing the body-fat
analysis in Section~\ref{sec:fat_analysis} and its associated figures,
is available from the
\href{https://github.com/heogden/adastrumm}
{\texttt{adastrumm} GitHub repository}. Code and saved results for
reproducing the simulation study and associated figures are available
from the
\href{https://github.com/heogden/simpleclustsim}
{\texttt{simpleclustsim} GitHub repository}.

\section*{Declaration of the use of generative AI and AI-assisted
technologies}

During the preparation of this work the author used ChatGPT to assist
with language editing and manuscript restructuring, writing and
checking parts of the simulation code, and refining regularity
conditions and technical details of proofs. After using this tool, the
author reviewed and edited the content as necessary and takes full
responsibility for the content of the publication.

\section*{Acknowledgement}

The author acknowledges the use of the IRIDIS High Performance
Computing Facility and associated support services at the University
of Southampton.

\bibliography{refs}
\bibliographystyle{plainnat}

\clearpage
\appendix
\renewcommand{\theequation}{S\arabic{equation}}
\section{Proofs}
\label{app:proofs}

\subsection{Proof of Theorem~\ref{thm:MLE}}

Write
\[
  Z_i=(n_i,x_i,Y_i),
  \qquad
  m_\theta(Z_i)=\ell_i(\theta;Y_i\mid n_i,x_i),
\]
and let \(E_0\) denote expectation under \(\theta_0^*\). For a local
working parameter \(\psi\), write
\(m_\psi=m_{T_a(\psi)}\). We first record the required likelihood
regularity.

\begin{lemma}
\label{lemma:likelihood_regularity}
Under Assumptions~\ref{ass:n_i}--\ref{ass:not_boundary}, part~(i)
below holds. If Assumption~\ref{ass:non_singular} also holds, then
part~(ii) holds.
\begin{enumerate}
\item[(i)] The map \(\theta\mapsto m_\theta(z)\) is continuous on
\(\Theta_c^*\) for each fixed \(z\). Moreover, for integrable random
variables \(F\) and \(L\),
\[
  \sup_{\theta\in\Theta_c^*}|m_\theta(Z_i)|\leq F(Z_i),
\]
and
\[
  |m_{\theta_1}(Z_i)-m_{\theta_2}(Z_i)|
  \leq L(Z_i)\|\theta_1-\theta_2\|,
  \qquad \theta_1,\theta_2\in\Theta_c^*.
\]
\item[(ii)] There is an open neighbourhood \(N\) of \(\psi_0^*\) on
which \(T_a\) is smooth and one-to-one, the score
\(\dot m_\psi=\nabla_\psi m_\psi\) is well defined,
\[
  E_0\|\dot m_{\psi_0^*}(Z_i)\|^2<\infty,
\]
and, for a square-integrable random variable \(G_N\),
\[
  \|\dot m_{\psi_1}(Z_i)-\dot m_{\psi_2}(Z_i)\|
  \leq G_N(Z_i)\|\psi_1-\psi_2\|,
  \qquad \psi_1,\psi_2\in N.
\]
\end{enumerate}
\end{lemma}

\begin{proof}
For \(\theta=(\beta_0,\beta,\eta)\), the conditional cluster
log-likelihood is
\[
  m_\theta(z_i)
  =
  -\frac{n_i}{2}\log(2\pi)
  -\frac12\log|\Sigma_i(\theta)|
  -\frac12\{y_i-m_i(\theta)\}^\T
  \Sigma_i(\theta)^{-1}
  \{y_i-m_i(\theta)\}.
\]
The basis functions are bounded on the compact interval \(\mathcal I\),
\(n_i\leq n_{\max}\), and \(\Theta_c^*\) is compact. Hence all
coefficients are bounded and \(\exp(2\eta)\) is uniformly bounded away
from zero. The mean, covariance, inverse covariance and their first two
derivatives are therefore uniformly bounded on the relevant parameter
sets. It follows that the likelihood and its derivatives are bounded by
constants times \(1+\|Y_i\|^2\). Normality and bounded cluster size give
the required integrable and square-integrable envelopes. Part (i)
then follows from the mean value theorem. If
Assumption~\ref{ass:non_singular} also holds,
Assumption~\ref{ass:not_boundary} allows a neighbourhood \(N\) to be
chosen within the regular canonical region, on which \(T_a\) is a
smooth one-to-one local parameterization. Applying the same bounds to
the score and Hessian proves part (ii).
\end{proof}

\begin{lemma}
\label{lemma:information_nonsingular}
Under Assumptions~\ref{ass:n_i}--\ref{ass:non_singular},
\(I_a(\psi_0^*)\) is positive definite.
\end{lemma}

\begin{proof}
Let \(v\) be a direction in the working parameter and write
\[
  (\dot\beta_0,\dot B,\dot\eta)
  =D T_a(\psi_0^*)v,
  \qquad
  B_0=B(\theta_0^*)=(\beta_1^0,\ldots,\beta_{K_0}^0),
\]
with
\[
  \dot Q=\dot B B_0^\T+B_0\dot B^\T.
\]
The directional derivatives of the conditional mean and covariance are
\[
  \dot m_i=X^{(i)}\dot\beta_0,
  \qquad
  \dot\Sigma_i
  =X^{(i)}\dot Q\{X^{(i)}\}^\T
   +2\sigma_0^2\dot\eta I_{n_i}.
\]
For a normal model,
\[
 v^\T I_a(\psi_0^*)v
 =E_0\!\left[
   \dot m_i^\T\Sigma_{0i}^{-1}\dot m_i
   +\frac12\operatorname{tr}
    \{\Sigma_{0i}^{-1}\dot\Sigma_i
      \Sigma_{0i}^{-1}\dot\Sigma_i\}
 \right].
\]
If this is zero, both derivatives vanish almost surely. Full support of
\(P_X\), continuity and orthonormality of the basis then give
\(\dot\beta_0=0\). On clusters with \(n_i\geq2\), the off-diagonal
entries give
\(b(x)^\T\dot Q b(x')=0\) throughout
\(\mathcal I\times\mathcal I\), and double integration against
\(b_r(x)b_s(x')\) gives \(\dot Q=0\). The diagonal entries then give
\(\dot\eta=0\).

It remains to identify \(\dot B\). Let
\(D_0=B_0^\T B_0\), whose diagonal entries are positive and distinct.
From \(\dot Q=0\), projection onto the orthogonal complement of the
column space of \(B_0\) gives \(\dot B=B_0A\) for some matrix \(A\),
and then \(A+A^\T=0\). Differentiating the orthogonality constraints
shows that the off-diagonal entries of
\(A^\T D_0+D_0A\) vanish. Thus, for \(j\ne k\),
\[
  \{(D_0)_{jj}-(D_0)_{kk}\}A_{jk}=0.
\]
The diagonal entries of \(D_0\) are distinct, so \(A=0\), and hence
\(\dot B=0\). Assumption~\ref{ass:non_singular} makes
\(D T_a(\psi_0^*)\) injective, so \(v=0\). Therefore the information
matrix is positive definite.
\end{proof}

\begin{proof}
We first prove consistency by applying Theorem~5.7 of
\citet{vanDerVaart1998}. Define
\[
  M_d(\theta)=\frac1d\sum_{i=1}^d m_\theta(Z_i),
  \qquad
  M(\theta)=E_0\{m_\theta(Z_i)\},
\]
and
\[
  q(\theta)=\frac{\gamma}{2\sigma(\theta)^2}w_E(\theta),
  \qquad
  M_{p,d}(\theta)=M_d(\theta)-d^{-1}q(\theta).
\]
Lemma~\ref{lemma:likelihood_regularity}(i) and compactness give
\[
  \sup_{\theta\in\Theta_c^*}|M_d(\theta)-M(\theta)|
  \xrightarrow{p}0.
\]
The continuous penalty is bounded on \(\Theta_c^*\), so the same
uniform convergence holds with \(M_d\) replaced by \(M_{p,d}\).

Correct specification gives
\[
  M(\theta)-M(\theta_0^*)
  =-E_0\!\left[
    \operatorname{KL}\{p_{\theta_0^*}(\cdot\mid n_i,x_i),
                 p_\theta(\cdot\mid n_i,x_i)\}
  \right]\leq0.
\]
Equality requires the conditional normal means and covariances to agree
almost surely. Mean equality holds \(P_X\)-almost surely and hence,
by continuity and full interval support, throughout \(\mathcal I\);
orthonormality then identifies \(\beta_0\). On clusters with
\(n_i\geq2\), equality of off-diagonal covariance entries similarly
identifies \(C(x,x')\) throughout
\(\mathcal I\times\mathcal I\). Writing
\(Q=BB^\T\), orthonormality gives
\[
  Q_{rs}
  =\int_{\mathcal I}\!\int_{\mathcal I}
    b_r(x)C(x,x')b_s(x')\,dx\,dx',
\]
so \(Q\) is identified. The diagonal covariance entries then identify
\(\sigma^2\). Because the columns of \(B\) are orthogonal with
positive distinct norms, the eigendecomposition of \(Q\) identifies
them up to sign and order; the canonical convention removes these
ambiguities. Thus \(M\) has the unique maximizer \(\theta_0^*\) on
\(\Theta_c^*\). Continuity and compactness give the separation condition
of Theorem~5.7, and therefore
\[
  \hat\theta^*\xrightarrow{p}\theta_0^*.
\]

We now prove asymptotic normality. By
Assumption~\ref{ass:non_singular}, choose a neighbourhood \(N\) of
\(\psi_0^*\) on which \(T_a\) is a smooth one-to-one local
parameterization, with image contained in the relative interior of
\(\Theta_c^*\). Consistency implies that, with probability tending to
one, \(\hat\theta^*\) lies in this image; define
\(\hat\psi^*=T_a^{-1}(\hat\theta^*)\) there, and extend it arbitrarily
outside this event.

For \(\psi\in N\), let
\[
  \varphi_\psi(Z_i)=\nabla_\psi m_\psi(Z_i),
  \qquad
  P_d\varphi_\psi=d^{-1}\sum_{i=1}^d\varphi_\psi(Z_i).
\]
Lemma~\ref{lemma:likelihood_regularity}(ii) gives the moment and local
Lipschitz conditions in Theorem~5.21 of
\citet{vanDerVaart1998}. Correct specification gives
\(E_0\{\varphi_{\psi_0^*}(Z_i)\}=0\), and differentiation under the
expectation gives
\[
  \left.\nabla_\psi E_0\{\varphi_\psi(Z_i)\}
  \right|_{\psi=\psi_0^*}
  =-I_a(\psi_0^*),
\]
which is nonsingular by
Lemma~\ref{lemma:information_nonsingular}.

On the event above, \(\hat\theta^*\) is an interior maximizer and the
first-order condition for the penalized criterion gives
\[
  P_d\varphi_{\hat\psi^*}
  =d^{-1}\nabla_\psi q\{T_a(\hat\psi^*)\}
  =O_p(d^{-1})=o_p(d^{-1/2}),
\]
since the penalty derivative is bounded locally. Theorem~5.21 therefore
gives
\[
  d^{1/2}(\hat\psi^*-\psi_0^*)
  =I_a(\psi_0^*)^{-1}
    d^{-1/2}\sum_{i=1}^d
    \varphi_{\psi_0^*}(Z_i)+o_p(1).
\]
The multivariate central limit theorem and the information identity now
give
\[
  d^{1/2}(\hat\psi^*-\psi_0^*)
  \xrightarrow{d}
  N\{0,I_a(\psi_0^*)^{-1}\},
\]
as required.
\end{proof}

\subsection{Proof of Theorem~\ref{thm:pop}}

\begin{proof}
Fix \(x,x'\in\mathcal I\). Consistency of \(\hat\theta^*\), established
in Theorem~\ref{thm:MLE} under
Assumptions~\ref{ass:n_i}--\ref{ass:not_boundary}, and continuity of
the maps from the model parameter to \(m(x)\) and \(C(x,x')\) give
\[
  \hat m(x)\xrightarrow{p}m(x),
  \qquad
  \hat C(x,x')\xrightarrow{p}C(x,x').
\]

Now suppose that Assumption~\ref{ass:non_singular} also holds. Then
\(T_a\) is smooth in a neighbourhood of \(\psi_0^*\). Since the spline
basis is fixed and finite-dimensional, the maps
\[
  g_m^x(\psi)=m\{x;T_a(\psi)\},
  \qquad
  g_C^{x,x'}(\psi)=C\{x,x';T_a(\psi)\},
\]
are continuously differentiable in that neighbourhood. Let
\[
  \dot g_m^x(\psi_0^*)
  =
  \left.
  \nabla_\psi g_m^x(\psi)
  \right|_{\psi=\psi_0^*},
  \qquad
  \dot g_C^{x,x'}(\psi_0^*)
  =
  \left.
  \nabla_\psi g_C^{x,x'}(\psi)
  \right|_{\psi=\psi_0^*}.
\]
Define
\begin{equation}
\label{eq:tau_m_def}
  \tau_m^2(x)
  =
  \dot g_m^x(\psi_0^*)^\T
  I_a(\psi_0^*)^{-1}
  \dot g_m^x(\psi_0^*),
\end{equation}
and
\begin{equation}
\label{eq:tau_C_def}
  \tau_C^2(x,x')
  =
  \dot g_C^{x,x'}(\psi_0^*)^\T
  I_a(\psi_0^*)^{-1}
  \dot g_C^{x,x'}(\psi_0^*).
\end{equation}

On the event, whose probability tends to one, on which
\(\hat\theta^*=T_a(\hat\psi^*)\) in the regular local
parameterization, the delta method applied to the asymptotic normality
part of Theorem~\ref{thm:MLE} gives
\[
  d^{1/2}\{\hat m(x)-m(x)\}
  \xrightarrow{d}
  N\{0,\tau_m^2(x)\},
\]
and
\[
  d^{1/2}\{\hat C(x,x')-C(x,x')\}
  \xrightarrow{d}
  N\{0,\tau_C^2(x,x')\}.
\]
\end{proof}

\subsection{Proof of Theorem~\ref{thm:traj}}

\begin{proof}
Fix \(i\), \(x\in\mathcal I\), and a possible cluster size \(n\) with
\(\operatorname{pr}(n_i=n)>0\). Conditional on \(n_i=n\), define
\[
  h_n^x(\theta,y,z)
  =
  E_\theta
  \{\mu_i(x)\mid Y_i=y,x_i=z,n_i=n\}.
\]
The normal conditioning formula shows that \(h_n^x\) is continuous
in \((\theta,y,z)\): the mean and covariance are continuous in
\(\theta\) and \(z\), and the covariance matrix of \(Y_i\) is
invertible because \(\sigma^2=\exp(2\eta)>0\).

The consistency part of Theorem~\ref{thm:MLE} gives
\(\hat\theta^*\xrightarrow{p}\theta_0^*\). Since \(\operatorname{pr}(n_i=n)>0\), the
same convergence holds conditional on \(n_i=n\). Conditional on this
event, Slutsky's theorem therefore gives
\[
  (\hat\theta^*,Y_i,x_i)
  \xrightarrow{d}
  (\theta_0^*,Y_i,x_i).
\]
The continuous mapping theorem then gives
\[
  h_n^x(\hat\theta^*,Y_i,x_i)
  -
  h_n^x(\theta_0^*,Y_i,x_i)
  \xrightarrow{p}0,
\]
conditional on \(n_i=n\). Since \(n_i\) has finite support, this also
holds unconditionally. The two terms are respectively
\(\hat\mu_i(x)\) and \(\mu_i^{\mathrm{or}}(x)\), proving the result.
\end{proof}

\section{Constructing a parameterization of the orthogonality-constrained component space}
\label{sec:construct_param}

We describe the set of working parameterizations used to enforce the
orthogonality constraint on the component coefficient vectors
\(\beta_1,\ldots,\beta_K\).  
We construct maps from unconstrained vectors
\[
  \alpha=(\alpha_1,\ldots,\alpha_K),
  \qquad
  \alpha_k\in\mathbb R^{n_B-k+1},
\]
to elements of \(\mathcal B_\perp\), the space of orthogonal $\beta$
components defined in \eqref{eq:B_perp}.  The full working parameter
is then
\[
  \psi=(\beta_0,\alpha,\eta),
  \qquad \eta=\log\sigma,
\]
where \(\beta_0\) and \(\eta\) are left unchanged by the transformation,
and only \(\alpha\) is used to construct
\((\beta_1,\ldots,\beta_K)\).

The construction is based on a sequence of reduced Householder
transformations.  For an integer \(m\geq 1\) and an index
\(a\in\{1,\ldots,m\}\), let \(e_a^{(m)}\) denote the \(a\)th standard
basis vector in \(\mathbb R^m\).  For a nonzero vector
\(x\in\mathbb R^m\), define
\[
  s_a(x)=
  \begin{cases}
  1, & x_a\geq 0,\\
  -1, & x_a<0,
  \end{cases}
  \qquad
  v_a(x)=x+s_a(x)\|x\|e_a^{(m)}.
\]
The associated Householder reflection is
\[
  H_a(x)
  =
  I_m - 2\frac{v_a(x)v_a(x)^\T}{v_a(x)^\T v_a(x)}.
\]
Then \(H_a(x)\) is orthogonal and symmetric, and
\[
  H_a(x)x = -s_a(x)\|x\|e_a^{(m)}.
\]
The sign choice in \(v_a(x)\) is a standard numerically stable
Householder choice \citep[see][section 3.2]{Watkins2002}, avoiding
cancellation when \(x\) is close to a multiple of \(e_a^{(m)}\).

We use a reduced version of \(H_a(x)\) to construct a basis for the
orthogonal complement of \(x\). Let \(E_{a,m}\) be
the \(m\times(m-1)\) matrix obtained by taking the identity
matrix \(I_{m-1}\) and inserting a row of zeroes as the \(a\)th
row. Then for \(z\in\mathbb R^{m-1}\), \(E_{a,m}z\) is the vector
in \(\mathbb R^m\) whose \(a\)th component is zero, and whose
remaining components are the entries of \(z\), in order.

Define
\[
  H_a^*(x) = H_a(x)E_{a,m}.
\]
The columns of \(H_a^*(x)\) are orthonormal and are orthogonal to \(x\):
if \(z\in\mathbb R^{m-1}\), then
\[
  x^\T H_a^*(x)z
  =
  \{H_a(x)x\}^\T E_{a,m}z
  =
  -s_a(x)\|x\| \{e_a^{(m)}\}^\T E_{a,m}z
  =
  0.
\]
Thus \(H_a^*(x)\) gives an orthonormal basis for the orthogonal
complement of \(x\) in \(\mathbb R^m\).

We now define the parameterization used to take an unconstrained
$\alpha$ to $\beta \in \mathcal{B}_\perp$. We consider a range of
possible parameterizations with this property, indexed by
\[
  a\in\mathcal A=\{1,\ldots,n_B-K+1\}.
\]
The restriction \(a\leq n_B-K+1\) ensures that the \(a\)th coordinate
exists in each reduced space used in the construction.

For some fixed index \(a\in\mathcal A\), define
\[
  T_{a,0}=I_{n_B},
  \qquad
  T_{a,k}
  =
  H_a^*(\alpha_1)H_a^*(\alpha_2)\cdots H_a^*(\alpha_k),
  \qquad k=1,\ldots,K-1.
\]

Then, for $k = 1, \ldots, K$, we let
\[
  \beta_k = T_{a,k-1}\alpha_k.
\]

We now check why this construction ensures orthogonality, so that
$\beta \in \mathcal{B}_\perp$. We have
\[
  T_{a,k}
  =
  T_{a,k-1}H_a^*(\alpha_k).
\]
This recursion ensures orthogonality.  Since the columns of
\(H_a^*(\alpha_k)\) span the orthogonal complement of \(\alpha_k\),
the columns of \(T_{a,k}\) span the orthogonal complement of
\(\operatorname{span}(\beta_1,\ldots,\beta_k)\).  Hence, for
\(j>k\),
\[
  \beta_j\in \operatorname{col}(T_{a,k}),
\]
whereas \(\beta_k=T_{a,k-1}\alpha_k\) is orthogonal to
\(\operatorname{col}(T_{a,k})\).  It follows that
\[
  \beta_j^\T\beta_k=0,\qquad j\ne k.
\]

The inverse map is also straightforward away from singularities.  Given
orthogonal nonzero vectors \(\beta_1,\ldots,\beta_K\), set
\[
  \alpha_1=\beta_1.
\]
Having recovered \(\alpha_1,\ldots,\alpha_{k-1}\), and hence
\(T_{a,k-1}\), the vector \(\beta_k\) lies in the column space of
\(T_{a,k-1}\).  We therefore recover
\[
  \alpha_k = T_{a,k-1}^\T\beta_k.
\]
Thus each fixed \(a\) gives a local parameterization of the
orthogonality-constrained component space.

The parameterization is not globally regular.  The transformation used
to construct \(H_a(\alpha_k)\) is not differentiable when the selected
coordinate \((\alpha_k)_a\) is zero, because the sign \(s_a(\alpha_k)\)
changes there. The construction also degenerates if \(\alpha_k=0\) for some
\(k<K\), because the subsequent orthogonal complement is then not
defined by the Householder formula.  Hence, for the
$\psi$-parameterization indexed by \(a\), we define a singular set
\begin{equation}
  S_a
  =
  \left\{
    (\beta_0,\alpha,\eta):
    \exists k\in\{1,\ldots,K-1\}
    \text{ such that }
    \alpha_k=0 \ \lor\  (\alpha_k)_a=0
  \right\}.
  \label{eq:singular_set}
\end{equation}
Only the first \(K-1\) blocks appear in this singularity condition,
because \(\alpha_K\) is not used to construct a subsequent orthogonal
complement.

These singularities are artefacts of the working parameterization, not
of the model.  A point in \(\mathcal B_\perp\) which is close to the
singular set for one value of \(a\) may be far from the singular set for
another value of \(a\).  This motivates using the collection of
parameterizations indexed by \(a\in\mathcal A\), rather than a single
fixed parameterization.  The practical choice of parameterization during
fitting and uncertainty calculation is described in
Section~\ref{sec:choose_param}.

\section{Choosing the working parameterization in the fitting procedure}
\label{sec:choose_param}

For any fixed \(a\in\mathcal A\), maximization of the penalized
log-likelihood can be carried out over the unconstrained working
parameter \(\psi=(\beta_0,\alpha,\eta)\), using the transformation
defined in Section~\ref{sec:construct_param}. In exact arithmetic, the
choice of \(a\) should not affect the estimated
\(\theta=(\beta_0,\beta,\eta)\), since each parameterization describes
the same constrained space, away from its singular set.

However, we could encounter two possible problems in a fixed
parameterization. First, optimization may be less numerically stable
if the parameter estimate is close to the singular set for a
particular choice of \(a\). Second, confidence intervals rely on a
local normal approximation. The quality of this approximation depends
on how well the penalized log-likelihood surface may be approximated
by a quadratic function around \(\hat\psi\). This approximation may be
poor if \(\hat\psi\) is close (relative to our level of uncertainty
about $\psi$) to the singular set for the chosen value of \(a\).

We therefore start the model fitting process with \(a=1\), and switch
parameterization in cases where we detect that the current choice of
\(a\) may cause either of those problems.

For a fitted value \(\hat\alpha\), define the point diagnostic
\[
  D_{\mathrm{pt}}(\hat\alpha,a)
  =
  \min_{k=1,\ldots,K-1}
  \frac{|(\hat\alpha_k)_a|}{\|\hat\alpha_k\|}.
\]
Large values of \(D_{\mathrm{pt}}\) indicate that the selected coordinate
is well away from zero in each block used to construct a subsequent
orthogonal complement. Small values indicate proximity to the singular
set for the current parameterization. If the Hessian of the penalized
log-likelihood is not negative definite, or if \(D_{\mathrm{pt}}\) is
below a small numerical threshold, taken to be \(10^{-5}\) by default,
we re-express the current fitted \(\beta\) under each candidate
parameterization \(a\in\mathcal A\), choose the value of \(a\) with
the largest point diagnostic, and repeat the optimization using this
re-expressed point as the starting value.

A separate diagnostic is used for the uncertainty calculation. Let
\(V\) denote the estimated covariance matrix for \(\hat\psi\), obtained
from the inverse negative Hessian. For the current parameterization
\(a\), define
\[
  D_{\mathrm{ci}}(\hat\psi,V,a)
  =
  \min_{k=1,\ldots,K-1}
  \frac{|(\hat\alpha_k)_a|}
       {\{V_{ka,ka}\}^{1/2}},
\]
where \(V_{ka,ka}\) denotes the diagonal entry of \(V\) corresponding to
the selected coordinate \((\alpha_k)_a\). This diagnostic is the minimum
number of standard errors between the fitted selected coordinate and the
nearest singularity \((\alpha_k)_a=0\). If \(D_{\mathrm{ci}}\) is large,
we use the current parameterization for the confidence interval
calculation.

If \(D_{\mathrm{ci}}\) is small (by default, less than 2), we consider
alternative parameterizations. Recomputing the Hessian for every
\(a\in\mathcal A\) would be unnecessarily expensive, so we first use a
local linear approximation to screen the candidates. Let
\[
  J_a(\hat\alpha_a)
  =
  \frac{\partial\,\operatorname{vec}\{\beta_a(\alpha)\}}
       {\partial \alpha^\T}
  \bigg|_{\alpha=\hat\alpha_a},
\]
be the Jacobian of the map from the working component parameters to the
orthogonal coefficient vectors, evaluated in the current
parameterization \(a\). For a candidate parameterization \(b\), let
\(\hat\alpha_b\) be the coordinates representing the same fitted
\(\hat\beta\) under parameterization \(b\), and define
\(J_b(\hat\alpha_b)\) analogously.

Small perturbations in the two working parameterizations should produce
the same first-order perturbation in \(\beta\). We therefore choose
\(A_{\alpha,b\leftarrow a}\) to solve 
\[
  J_b(\hat\alpha_b) A_{\alpha,b\leftarrow a} = 
  J_a(\hat\alpha_a).
\]
If \(J_b(\hat\alpha_b)\) has full column rank, this gives
\[
  A_{\alpha,b\leftarrow a}
  =
  \{J_b(\hat\alpha_b)^\T J_b(\hat\alpha_b)\}^{-1}
  J_b(\hat\alpha_b)^\T J_a(\hat\alpha_a).
\]
This is the linear approximation used to transform perturbations in the
current \(\alpha\)-coordinates to perturbations in the candidate
\(\alpha\)-coordinates.

The full working parameter is
\(\psi=(\beta_0,\alpha,\eta)\). Since \(\beta_0\) and \(\eta\) are
unchanged by switching the component parameterization, the corresponding
linear approximation for the full working parameter is
\[
  A_{\psi,b\leftarrow a}
  =
  \begin{pmatrix}
    I_{n_B} & 0 & 0 \\
    0 & A_{\alpha,b\leftarrow a} & 0 \\
    0 & 0 & 1
  \end{pmatrix}.
\]
If \(V_a\) is the covariance matrix for \(\hat\psi_a\), obtained from
the inverse negative Hessian in the current parameterization, then the
candidate covariance matrix is approximated by
\[
  V_b
  \approx
  A_{\psi,b\leftarrow a}
  V_a
  A_{\psi,b\leftarrow a}^\T .
\]
We compute \(D_{\mathrm{ci}}(\hat\psi_b,V_b,b)\) using this approximate
covariance matrix for each candidate parameterization \(b\).

We then take the candidate parameterization with the largest approximate
value of \(D_{\mathrm{ci}}\), recompute the Hessian exactly in that
parameterization, and switch to it only if the exact value of
\(D_{\mathrm{ci}}\) improves on the current one. This final switching
step does not change the fitted model: $\hat \theta$ is unchanged, and
only the working-coordinate representation is altered.  The purpose of
the switch is solely to obtain a more stable local normal
approximation for uncertainty calculation.

\section{Sampling basis coefficients for confidence intervals}
\label{sec:ci_details}

Section~\ref{sec:ci} describes how we find pointwise confidence
intervals for \(\mu_i(x)\), based on a simulated sample
\(\{\delta_i^{(1)},\ldots,\delta_i^{(n_S)}\}\) for the basis
coefficients \(\delta_i\) of the individual trajectories. Here we
describe how to find this sample.

The confidence interval calculation treats \(K\) and \(\gamma\) as fixed
at their chosen values. It also uses a fixed working parameterization,
chosen as described in Section~\ref{sec:choose_param}. Write
this working parameterization as
\[
  \theta = T_a(\psi),
  \qquad
  \psi = (\beta_0,\alpha,\eta),
\]
where \(\eta=\log\sigma\). Let \(\hat\psi_a\) be the fitted value in this
working parameterization, so that
\[
  \hat\theta = T_a(\hat\psi_a).
\]
We find the negative Hessian matrix
\[
  H_\psi
  =
  -\nabla_\psi^2 \ell_p\{T_a(\psi);\gamma,K\}
  \bigg|_{\psi=\hat\psi_a},
\]
and use
\[
  V_\psi = H_\psi^{-1},
\]
as an approximate covariance matrix for \(\hat\psi_a\).

For each sample \(j=1,\ldots,n_S\), we proceed as follows.

\begin{enumerate}
\item Generate a working-parameter draw
\[
  \psi^{(j)} \sim N_p(\hat\psi_a,V_\psi).
\]
Map this draw to the constrained model parameter by setting
\[
  \theta^{(j)} = T_a\{\psi^{(j)}\}.
\]
Write
\[
  \theta^{(j)}
  =
  \left(\beta_0^{(j)},\beta_1^{(j)},\ldots,\beta_K^{(j)},\eta^{(j)}\right),
  \qquad
  [\sigma^2]^{(j)} = \exp\{2\eta^{(j)}\}.
\]
For \(k=0,\ldots,K\), define
\[
  f_k^{(j)}(x) = \{\beta_k^{(j)}\}^\T b(x).
\]

\item For each cluster \(i=1,\ldots,d\), form the vectors
\[
  f_{ki}^{(j)}
  =
  \left\{
    f_k^{(j)}(x_{i1}),\ldots,f_k^{(j)}(x_{in_i})
  \right\}^\T,
  \qquad k=0,\ldots,K.
\]
Let \(F_i^{(j)}\) be the \(n_i\times K\) matrix with columns
\(f_{1i}^{(j)},\ldots,f_{Ki}^{(j)}\). The marginal covariance matrix of
\(Y_i\), conditional on \(\theta^{(j)}\), is
\[
  \Sigma_{Y,i}^{(j)}
  =
  [\sigma^2]^{(j)} I_{n_i}
  +
  F_i^{(j)}\{F_i^{(j)}\}^\T .
\]

\item Conditional on \(\theta^{(j)}\), \(y_i\) and \(x_i\), the random
effects have a normal distribution
\[
  u_i \mid y_i,x_i,\theta^{(j)}
  \sim
  N_K\left(m_{u,i}^{(j)}, V_{u,i}^{(j)}\right),
\]
where
\[
  m_{u,i}^{(j)}
  =
  \{F_i^{(j)}\}^\T
  \{\Sigma_{Y,i}^{(j)}\}^{-1}
  \left(y_i-f_{0i}^{(j)}\right),
\]
and
\[
  V_{u,i}^{(j)}
  =
  I_K
  -
  \{F_i^{(j)}\}^\T
  \{\Sigma_{Y,i}^{(j)}\}^{-1}
  F_i^{(j)}.
\]
Generate
\[
  u_i^{(j)}
  \sim
  N_K\left(m_{u,i}^{(j)}, V_{u,i}^{(j)}\right).
\]

\item Finally, define the sample of the basis coefficients for
the individual trajectory by
\[
  \delta_i^{(j)}
  =
  \beta_0^{(j)}
  +
  \sum_{k=1}^K u_{ik}^{(j)}\beta_k^{(j)}.
\]
\end{enumerate}

Repeating this process for \(j=1,\ldots,n_S\) gives the samples
\[
  \left\{
    \delta_i^{(1)},\ldots,\delta_i^{(n_S)}
  \right\},
  \qquad i=1,\ldots,d.
\]
These are the samples used in Section~\ref{sec:ci} to construct
pointwise confidence intervals for the individual trajectories and any
derived quantities.

\section{Details of choosing the number of components \(K\)}
\label{sec:choose_K_details}

For fixed \(\gamma\), let \(\hat\sigma_K^2\) denote the estimated
error variance under a model with \(K\) components. As \(K\) increases,
the fit approaches a limiting residual variance
\[
  \hat\sigma_\infty^2
  =
  \lim_{K\rightarrow\infty}\hat\sigma_K^2.
\]
The fraction of variance explained by \(K\) components is
\[
  \operatorname{FVE}(K)
  =
  \frac{\hat\sigma_0^2-\hat\sigma_K^2}
       {\hat\sigma_0^2-\hat\sigma_\infty^2}.
\]
We seek the smallest \(K\) for which
\(\operatorname{FVE}(K)>t_{\mathrm{FVE}}\).

Because \(\hat\sigma_\infty^2\) cannot be computed directly, we
approximate it by \(\hat\sigma_{K_{\max}}^2\), giving
\[
  \operatorname{FVE}(K;K_{\max})
  =
  \frac{\hat\sigma_0^2-\hat\sigma_K^2}
       {\hat\sigma_0^2-\hat\sigma_{K_{\max}}^2}.
\]
Starting with \(K_{\max}=2\), we fit all models with
\(K\leq K_{\max}\). If
\(\operatorname{FVE}(K_{\max}-1;K_{\max})>t_{\mathrm{FVE}}\), we
choose \(K=K_{\max}-1\). Otherwise, we increase \(K_{\max}\) by one,
fit the additional model and repeat. We use
\(t_{\mathrm{FVE}}=0.999\) throughout.

\section{Derivation of criterion for choosing the smoothing parameter $\gamma$}
\label{sec:gamma_choice_details}

We derive the approximate marginal likelihood criterion used to choose
$\gamma$. The derivation follows the usual Bayesian interpretation of
smoothing penalties \citep[Section~6.2.6]{Wood2017}, but with integration
over the orthogonality-constrained coefficient space. For fixed $\gamma$
and $K$, we treat $\sigma$ as fixed at its penalized maximum likelihood
estimate $\hat\sigma_{\gamma,K}$.

Write
\[
  \theta_c=(\beta_0,\beta),
  \qquad
  \beta=(\beta_1,\ldots,\beta_K),
\]
for the coefficient part of the parameter, and let
\[
  \Theta_{c,\perp}
  =
  \left\{
  \theta_c=(\beta_0,\beta):
  \beta\in\mathcal B_\perp
  \right\},
\]
be the orthogonality-constrained coefficient space. The penalty gives the
prior kernel
\[
  \pi_0(\theta_c)
  =
  \exp\left\{
  -\frac{\gamma}{2\hat\sigma_{\gamma,K}^2}
  \theta_c^\T R_K\theta_c
  \right\},
  \qquad
  \theta_c\in\Theta_{c,\perp},
\]
where
\[
  R_K=\operatorname{blockdiag}(S,\ldots,S),
\]
with $K+1$ copies of $S$. Since $R_K$ is rank deficient, this prior is
improper: it is normal in the penalized directions and flat in the null
directions of the penalty.

We separate the calculation into two parts. First, define
\[
  I_{\gamma,K}
  =
  \int_{\Theta_{c,\perp}}
  f(y\mid\theta_c)\pi_0(\theta_c)
  \,dV(\theta_c),
\]
where $dV(\theta_c)$ denotes the volume element induced on the constrained
coefficient space by the surrounding Euclidean coefficient space. Second,
we add the finite normalizing contribution from the penalized directions
of the improper prior.

To compute $I_{\gamma,K}$, use a local working parameterization
\[
  \psi_c=(\beta_0,\alpha),
  \qquad
  \theta_c(\psi_c)=\{\beta_0,\beta(\alpha)\}.
\]
Let
\[
  J(\psi_c)
  =
  \frac{\partial\theta_c(\psi_c)}{\partial\psi_c},
  \qquad
  G(\psi_c)=J(\psi_c)^\T J(\psi_c).
\]
In this parameterization,
\[
  dV(\theta_c)=|G(\psi_c)|^{1/2}\,d\psi_c.
\]
Therefore
\begin{equation}
  I_{\gamma,K}
  =
  \int
  f\{y\mid\theta_c(\psi_c)\}
  \exp\left[
  -\frac{\gamma}{2\hat\sigma_{\gamma,K}^2}
  \theta_c(\psi_c)^\T R_K\theta_c(\psi_c)
  \right]
  |G(\psi_c)|^{1/2}
  \,d\psi_c.
  \label{eq:integrated_likelihood_psi}
\end{equation}

We next approximate the finite normalizing contribution from the penalized
directions of the prior. Let $\hat\psi_c$ be the fitted value for the
current $\gamma$ and $K$, and write
\[
  u=\psi_c-\hat\psi_c.
\]
Using the local approximations
\[
  \theta_c(\psi_c)
  \simeq
  \hat\theta_c+\hat J u,
  \qquad
  G(\psi_c)\simeq \hat G,
\]
where
\[
  \hat J=J(\hat\psi_c),
  \qquad
  \hat G=G(\hat\psi_c)=\hat J^\T\hat J,
\]
the prior kernel is locally approximated, up to terms which do not affect
the normalizing constant, by
\[
  \exp\left\{
  -\frac{1}{2}u^\T P u
  \right\}
  |\hat G|^{1/2},
\]
where
\[
  P
  =
  \frac{\gamma}{\hat\sigma_{\gamma,K}^2}
  \hat J^\T R_K\hat J .
\]
This is still an improper normal approximation, because $P$ is rank
deficient. The normalizing constant is therefore computed only over the
penalized directions.

To express this normalizing contribution, let
\[
  \hat G=LL^\T,
  \qquad
  z=L^\T u.
\]
Then
\[
  |\hat G|^{1/2}\,du=dz,
\]
and
\[
  u^\T P u
  =
  z^\T L^{-1}PL^{-\T}z .
\]
Let $\lambda_1,\ldots,\lambda_r$ be the positive eigenvalues of
$L^{-1}PL^{-\T}$. Equivalently, these are the positive generalized
eigenvalues solving
\[
  Pv=\lambda\hat Gv.
\]
Write
\[
  |P|_{+,\hat G}
  =
  \prod_{j=1}^r\lambda_j,
  \qquad
  r=\operatorname{rank}_{\hat G}(P).
\]
The local approximation to the log normalizing contribution from the
penalized directions is therefore
\begin{equation}
  \log\tilde C_{\gamma,K}
  =
  \frac{1}{2}\log |P|_{+,\hat G}
  -
  \frac{r}{2}\log(2\pi).
  \label{eq:prior_norm_contribution}
\end{equation}

We now approximate the integral $I_{\gamma,K}$. Let $H$ be the negative
Hessian, with respect to $\psi_c$, of the penalized log-likelihood at
$\hat\psi_c$, with $\sigma$ held fixed at $\hat\sigma_{\gamma,K}$. For a
positive definite matrix $A$, write $|A|_{\hat G}$ for the product of all
generalized eigenvalues solving
\[
  Av=\lambda\hat Gv.
\]
Equivalently, if $\hat G=LL^\T$, then
\[
  |A|_{\hat G}=|L^{-1}AL^{-\T}|.
\]
A first-order Laplace approximation to \eqref{eq:integrated_likelihood_psi}
gives
\[
  \log\tilde I_{\gamma,K}
  =
  \ell_p(\hat\theta_{\gamma,K})
  +
  \frac{\dim(\psi_c)}{2}\log(2\pi)
  -
  \frac{1}{2}\log |H|_{\hat G}.
\]
Combining this Laplace approximation with the local normalizing
contribution from \eqref{eq:prior_norm_contribution} gives the
labelled-component criterion
\[
  \log\tilde m_{\gamma,K}^{\mathrm{lab}}
  =
  \log\tilde I_{\gamma,K}
  +
  \log\tilde C_{\gamma,K}.
\]
Hence
\[
  \log\tilde m_{\gamma,K}^{\mathrm{lab}}
  =
  \ell_p(\hat\theta_{\gamma,K})
  +
  \frac{1}{2}\log |P|_{+,\hat G}
  -
  \frac{1}{2}\log |H|_{\hat G}
  +
  \frac{M}{2}\log(2\pi),
\]
where
\[
  M
  =
  \dim(\psi_c)-r
  =
  \dim(\psi_c)-\operatorname{rank}_{\hat G}(P).
\]

Finally, the labelled orthogonal component space counts each canonical
model $K!2^K$ times, through permutations and sign changes of the
components. We therefore use
\[
  \log\tilde m_{\gamma,K}
  =
  \log\tilde m_{\gamma,K}^{\mathrm{lab}}
  -
  \log(K!2^K),
\]
or equivalently
\[
  \log\tilde m_{\gamma,K}
  =
  \ell_p(\hat\theta_{\gamma,K})
  +
  \frac{1}{2}\log |P|_{+,\hat G}
  -
  \frac{1}{2}\log |H|_{\hat G}
  +
  \frac{M}{2}\log(2\pi)
  -
  \log(K!2^K).
\]
This is the criterion in \eqref{eq:approx_laml_gamma}.

The exact integral $I_{\gamma,K}$ is independent of the particular
smooth working parameterization used to compute it. The approximation
above is also invariant to smooth changes of working coordinates in
exact arithmetic, because the determinant terms are written relative
to the metric $\hat G$. In finite precision computation, small
numerical differences may remain, especially when a working
parameterization is close to a singularity.

\section{Choice of methods for the simulation studies}
\label{sec:simstudies_pilot}

\subsection{Preliminary comparison and decision criteria}

Before running the full simulation studies, we conducted a smaller
preliminary comparison to choose tuning options and
population-level estimators, and to identify methods which were not
sufficiently competitive or reliable to take forward. For each of the
three data-generating processes described in
Section~\ref{sec:sim_generation}, we used two simulation runs at each
of the five settings
\[
(d,n_i) \in
\{(50,2),(50,10),(200,5),(500,2),(500,10)\}.
\]
These settings include the smallest and largest numbers of individuals
and the sparsest and densest trajectories used in the full simulation
study, together with the intermediate setting $(d,n_i)=(200,5)$.

To make errors comparable between settings, we expressed the RMISE for
individual trajectories relative to the corresponding individual-level
oracle bound, and the root mean squared 2-Wasserstein error for the
population process relative to the empirical population-level bound
defined in Section~\ref{sec:oracle_bounds}. We also considered
confidence-interval coverage and width, fitting and prediction
failures, numerical stability, and model-fitting time. We made one
global choice of tuning options and population-level estimator for
each method family, rather than selecting the best option separately
in each simulation setting.

We first chose the population-level estimator, then selected tuning
options within each method family, and finally decided whether each
method family should be retained for the full simulation study. The
preliminary comparison was used only for these global methodological
decisions; it was not used to determine the final set of
case-specific simulation runs.

After selecting the methods and tuning options, we carried out a
separate computational feasibility trial over all case and subcase
combinations intended for the full simulation study. For each selected
method and simulation setting, a small number of independent trial
runs were used to record model-fitting time, memory failures, timeouts
and fitting errors. These runs were used to identify settings in which
a method was systematically computationally infeasible, and to estimate
the resources required to allocate the full simulations across
parallel processes. The final case-by-setting run matrix and resource
allocation were fixed before examining the results from the full
simulation study. Occasional fitting failures for otherwise feasible
methods were retained as part of their assessed performance rather
than being used automatically to exclude the corresponding setting.

The Local-GAM method was retained only for settings with $n_i \geq 5$,
because fitting a separate smooth curve to each cluster is not a
meaningful baseline when each cluster has only two or three
observations. Local-GAM confidence intervals were also excluded from
the plotted interval comparisons, since they were extremely wide and
not informative for the comparisons of interest.

Complete results from the preliminary comparison, together with the
summaries used to make the global method choices, are provided as
processed CSV files in
the \href{https://github.com/heogden/simpleclustsim}
{\texttt{simpleclustsim} GitHub repository}. The repository also
includes processed CSV summaries from the computational feasibility
trial and the final run matrix for the full simulation studies.

\subsection{Implementation details for retained methods}
\label{sec:simstudies_implementation}

For Local-GAM, we fit a separate smooth spline to each cluster using
the \texttt{gam} function in \texttt{mgcv}. We retain this method only
for \(n_i\geq5\), since a separate smooth fit is not meaningful with
only two or three observations per cluster.

For HGAM-GS, we use the \texttt{bam} function in \texttt{mgcv} to fit
the hierarchical generalized additive model described by
\citet{Pedersen2019}. For PACE, we use the \texttt{FPCA} function in
the \texttt{fdapace} R package \citep{fdapace}, retaining components
that explain 99\% of the estimated variation, selecting bandwidths for
the mean and covariance smooths by generalized cross-validation, and
using default values for the remaining arguments. For FACE, we use
the \texttt{face.sparse} function in the \texttt{face} R package,
retaining components that explain 99\% of the estimated variation and
using default values for the remaining arguments.

For bayesFPCA, we use the \texttt{run\_vmp\_fpca} function in the
\texttt{bayesFPCA} R package. In the correctly specified two- and
three-component cases, we supply the true value of \(K\), so this
method is labelled as using oracle \(K\). We omit bayesFPCA from the
growth-curve case, where there is no true finite value of \(K\).

In the growth-curve case, we fit the correctly specified SITAR model
using the \texttt{sitar} R package \citep{Cole2010,sitar}.

\subsection{Choice of population-level estimator}
\label{sec:simstudies_gp_choice}

Section~\ref{sec:pop_inference} defines the plug-in, empirical and
empirical-corrected estimators of the population mean and covariance
functions.

For PACE and FACE, we additionally considered the smoothed mean and
covariance estimates returned directly by the fitting software, which
we call the \emph{direct} estimator. Unlike the plug-in estimator, the
direct estimator uses the full estimated covariance and does not
reflect the selected truncation to a particular number of components.

Within each method and preliminary simulation setting, we compared the
available population-level estimators using root mean squared
2-Wasserstein error relative to the empirical population-level bound.
We also calculated the ratio between each estimator's error and that of
the best available estimator in the same setting. We summarized the
relative errors across settings using geometric means, while also
considering worst-case performance. A single estimator was then chosen
for each method family.

For AdaStruMM, the plug-in and empirical-corrected estimators gave very
similar population-level errors. The plug-in estimator is obtained directly from the fitted AdaStruMM
model, while the empirical-corrected estimator estimates the population distribution of
trajectories by combining the fitted conditional mean trajectories
with the average conditional uncertainty in the random effects. In the
full simulation settings, the empirical-corrected estimator gave
slightly smaller root mean squared 2-Wasserstein error in most of the
correctly specified two- and three-component settings, while giving
very similar performance to the plug-in estimator in the growth-curve
setting. We therefore use the empirical-corrected estimator for
AdaStruMM population-level inference.

For bayesFPCA with the true number of components supplied, the
geometric mean relative population error was $1.54$ for the plug-in
estimator and $1.64$ for the empirical estimator. The plug-in estimator was within 10\% of the best estimator in every setting. The same conclusion
was supported by the variants which selected the number of components
by percentage of variation explained: their plug-in estimators were within 1.5\% of the best available
estimator in every setting. We therefore
use the plug-in estimator for bayesFPCA.

For PACE, the plug-in and direct estimators performed almost
identically on average, with geometric mean relative errors of $2.68$ and $2.67$,
respectively, while the empirical estimator gave $2.80$. The plug-in
estimator had better worst-case performance than the direct estimator
and, unlike the direct estimator, represents the covariance using the
selected components. We therefore use the plug-in estimator for PACE.

For FACE, the empirical estimator had a lower geometric mean relative
error than the plug-in estimator, but was substantially less stable:
its maximum relative error was $11.60$, compared with $5.32$ for the
plug-in estimator. The plug-in and direct estimators had similar
average and worst-case performance, but only the plug-in estimator
reflects the chosen percentage-of-variation truncation. We therefore
use the plug-in estimator for FACE.

Only the empirical estimator was available for Local-GAM, HGAM-GS and
the fitted SITAR model, and we use it for these methods. Thus, in the
full simulation study, we use the empirical-corrected
population-level estimator for AdaStruMM, the plug-in estimator for
PACE, FACE and bayesFPCA, and the empirical estimator for Local-GAM,
HGAM-GS and SITAR.

\subsection{Choices within method families}
\label{sec:simstudies_method_choices}

We next used the preliminary comparison to choose between the tuning
options considered within each retained method family. No corresponding
comparison was required for AdaStruMM, Local-GAM, HGAM-GS or the fitted
SITAR model, for which we used the specifications described in
Section~\ref{sec:sim_methods}.

\medskip
\noindent\textit{PACE: select $K$ by FVE99 and use GCV smoothing.} For PACE, we considered choosing the number of components $K$ by AIC,
BIC, or retaining sufficient components to explain 95\% or 99\% of the
estimated variation. In the two- and three-component cases, we also
considered supplying the true value of $K$. For each choice of $K$, we
compared the package's default bandwidth selection with generalized
cross-validation (GCV).

Across the 15 preliminary settings, the FVE99-GCV version had the
lowest geometric mean relative population error, at $2.68$, compared
with $2.72$ for both AIC-GCV and BIC-GCV and $2.96$ for FVE95-GCV. It
also had better worst-case population performance: its maximum relative
error was $6.61$, compared with $7.25$ for the AIC-GCV and BIC-GCV
versions. The variants using the default bandwidth choice generally
had larger and less stable population errors, with maximum relative
errors of up to $15.91$. Differences in individual-trajectory error
between the leading PACE variants were comparatively small.

In the ten preliminary settings for which the true finite value of
$K$ was available, FVE99-GCV and Oracle-GCV gave identical
individual- and population-level errors. Thus, supplying the true
number of components gave no improvement over the fully data-driven
FVE99 choice in these settings. We therefore use FVE99 to select $K$
and GCV to select the smoothing bandwidths in the full simulation
study.

\medskip
\noindent\textit{FACE: retain components explaining 99\% of variation.} For FACE, we compared retaining sufficient components to explain 95\%
or 99\% of the estimated variation, using the default values for the
remaining tuning parameters. Their overall population performance was
very similar: the geometric mean relative population errors were
$2.541$ and $2.538$ for the 95\% and 99\% versions, respectively.
However, the 99\% version had better worst-case performance, with a
maximum relative error of $5.32$ rather than $5.68$, and performed
more clearly better in the growth-curve case. Within that case, the
geometric mean relative population errors were $2.38$ and $2.81$ for
the 99\% and 95\% versions, respectively.

The FACE percentage-of-variation choice does not alter its fitted
individual trajectories: the prediction method uses the full smoothed
covariance estimate, whereas the selected number of components affects
the truncated FPC representation used for population-level inference.
The individual-trajectory errors and confidence intervals were
therefore identical for the two versions. We select the 99\% version
because of its slightly better overall and substantially better
growth-curve population performance.

\medskip
\noindent\textit{bayesFPCA: use the true $K$ as a favourable benchmark.} For bayesFPCA, we first fitted a model with a maximum of 15 components
and retained sufficient components to explain either 95\% or 99\% of
the fitted variation. In the two- and three-component cases, we also
considered supplying the true number of components. On the ten
preliminary settings from these two cases, the oracle-$K$ version gave
a substantial improvement. Its geometric mean relative population
error was $1.54$, compared with $2.97$ for PVE99 and $3.32$ for PVE95.
The corresponding geometric mean relative individual-trajectory
errors were $3.16$, $5.04$ and $5.84$, respectively.

We also investigated the model-choice procedure supplied with
bayesFPCA. Because this procedure was extremely computationally
expensive, we ran it only for two replicates of the smallest
two-component setting. Its results were close to those from the
oracle-$K$ version, but each fit took approximately 30 minutes even in
this small setting. Applying this procedure throughout the full
simulation design would therefore have been impractical.

We consequently retain bayesFPCA with the true value of $K$ in the
two- and three-component cases. This is not a practical data-driven
procedure, and we label it explicitly as ``bayesFPCA (oracle $K$)''.
Its role is to provide a favourable assessment of the performance that
bayesFPCA might achieve if the difficulty of selecting the number of
components were resolved. We do not include bayesFPCA in the
growth-curve simulations, where the data-generating process has no
true finite value of $K$.

\subsection{Methods not taken forward}
\label{sec:simstudies_excluded_methods}

We also included two methods implemented in the \texttt{refund} R
package \citep{refund} in the preliminary comparison, but did not take
either forward to the full simulation study. For both methods, only
the empirical population-level estimator was available. We therefore
based the decision primarily on individual-trajectory accuracy,
confidence interval performance and computational reliability, rather
than on population-level error.

\medskip
\noindent\textit{The method of \cite{Di2009} was not retained.}
We fitted this method using the \texttt{fpca.sc} function, choosing $K$
 to explain either 95\%
or 99\% of the estimated variation. In the two- and three-component
cases, we also considered supplying the true number of components.

At least one of the 95\% and 99\% variants produced finite trajectory
estimates in 11 of the 15 preliminary settings. To give the method a
favourable comparison, we considered the better of
these two variants in each available setting. This version had lower
RMISE than the selected PACE method in six of the
11 settings, but lower RMISE than the selected FACE method in only one.
It outperformed both methods in only the smallest two-component
setting, $(d,n_i)=(50,2)$, and that comparison was based on only one
successful preliminary run.

The method also showed substantial instability. Three settings
exceeded the 8~GB memory limit, and all fits failed in another
setting. The reported confidence intervals had zero average width in
11 of the 21 finite variant-by-setting results, and one 99\% fit gave
an individual-trajectory error more than 1000 times the corresponding
oracle bound. Supplying the true number of components did not provide a
consistent improvement. We therefore did not retain this method.

\medskip
\noindent\textit{The method of \cite{Goldsmith2013} was not retained.}
We fitted this method using the \texttt{ccb.fpc} function, again
choosing $K$ to explain either 95\%
or 99\% of the estimated variation. The method completed within the
memory limit in only seven of the 15 preliminary settings; all eight
remaining settings exceeded the 8~GB memory limit.

Among the seven settings with finite results, the better of the two
variants had lower RMISE than the selected PACE
method in two settings and than the selected FACE method in one, and
outperformed both in only one setting. Performance was also unstable
between the two variants. In some completed settings, relative
RMISE was as large as 69, and average confidence
interval widths exceeded 60. Thus the method offered no consistent
accuracy advantage over the retained functional principal components
methods, while being substantially less computationally reliable. We
therefore did not include it in the full simulation study.

\subsection{Additional evaluation details}
\label{sec:sim_evaluation_details}

Occasionally, a fitted method or a downstream summary was unavailable
for a small number of simulated datasets. For any metric shown in the
figures, if it was available in at least 95\% of simulation runs, the
plotted estimate and uncertainty band were computed from the subset of
available runs. Method--metric combinations with larger failure rates
were omitted from the relevant plots and discussed separately.

The Local-GAM confidence intervals were extremely wide and
uninformative in these settings, reflecting the fact that each
trajectory was estimated from that cluster's data alone. We therefore
omit Local-GAM confidence-interval coverage and width from the figures.
Local-GAM remains in the trajectory-error and timing comparisons for
\(n_i\geq5\) as a deliberately simple local baseline.

\section{Comparing population-level inference in simulations}
\label{sec:sim_comparisons_pop}

In the two- and three-component cases, model~\eqref{eq:model} is
correct for some $K = K_0$, so we use \eqref{eq:cov_fpc} to find
the covariance function. In the growth curve case,
model \eqref{eq:model} is not correct for any finite $K$, and the
population-level process is not a Gaussian process. Nonetheless, we
can still compute the true mean and covariance functions, and we study
the error in the estimated Gaussian process relative to a Gaussian
process with these true mean and covariance functions.  The
notation \(\hat m\) and \(\hat C\) below denotes the chosen
population-level estimator for the method being evaluated, as
specified in Section~\ref{sec:simstudies_gp_choice}.

To compare the reference and estimated Gaussian processes, we use the
squared 2-Wasserstein distance. We first discretize these processes by
considering all functions evaluated on
a fine grid of time points $(t_1, \ldots, t_N)^\T$. Write $\mu_i =
(\mu_i(t_1), \ldots, \mu_i(t_N))^\T$, $m$ for the $N$-vector with
entries $m_j = m(t_j)$ and $C$ for the $N \times N$ matrix with
entries $C_{j_1, j_2} = C(t_{j_1}, t_{j_2})$, and similarly $\hat m$
and $\hat C$ are the discretized versions of the fitted mean and
covariance functions.  So, the reference Gaussian process is discretized to
$\mu_i \sim N_N(m, C)$, and we estimate $\mu_i \sim N_N(\hat m, \hat
C)$.  The squared 2-Wasserstein distance between $N_N(m, C)$ and
$N_N(\hat m, \hat C)$ is \citep{Olkin1982}
\[
  W_2^2(N)
  =
  \|\hat m-m\|_2^2
  +
  \operatorname{tr}
  \left\{
    \hat C+C-2(C^{1/2}\hat C C^{1/2})^{1/2}
  \right\}.
\]
We average this to give $\bar W_2^2 = N^{-1} W_2^2(N)$, which stabilizes
in the large-$N$ limit. In simulation studies, we use a regular grid
of $N = 100$ time points between $t_{\min}$ and $t_{\max}$ to
calculate a squared 2-Wasserstein error $\bar W_2^{2, (i)}$ for each simulation
run ($i = 1, \ldots, n_\text{sim}$). We aggregate this to give the
mean squared 2-Wasserstein error (MWE) across simulation runs,
\[\operatorname{MWE} = \frac{1}{n_\text{sim}} \sum_{i=1}^{n_\text{sim}} \bar W_2^{2, (i)},\]
where $n_\text{sim} = 100$, and report the root mean squared 2-Wasserstein
error, $\operatorname{RMWE} = \operatorname{MWE}^{1/2}$.

\section{Additional three-component simulation}
\label{sec:three_component_simulation}

We consider an additional correctly specified mixed-effects model with
\(K=3\) components,
\[
  \mu_i(x)
  =
  h(x)+u_{i1}h(x)+u_{i2}+u_{i3}\cos(x),
  \qquad
  h(x)=x/2+\sin(x),
\]
where \(u_{i1}\), \(u_{i2}\) and \(u_{i3}\) are independent
\(N(0,1)\) random variables. We generate \(x_{ij}\) uniformly on
\([0,3\pi]\), with measurement-error standard deviation
\(\sigma=0.1\).

The remaining simulation design, including the values of \(d\) and
\(n_i\), the 100 replicates, the methods, oracle benchmarks and
evaluation criteria, is as described in
Sections~\ref{sec:sim_generation}--\ref{sec:sim_comparisons}.

\begin{figure}
  \begin{subfigure}{\textwidth}
    \centering
    \resizebox{0.96\linewidth}{!}{\input{figures/RMISE-3re.tex}}
    \caption{Root mean integrated squared error
      in individual trajectories.}\label{fig:RMISE-3re}
  \end{subfigure}
 \begin{subfigure}{\textwidth}
    \centering
    \resizebox{0.96\linewidth}{!}{\input{figures/coverage-3re.tex}}
    \caption{Coverage of confidence
      intervals for individual trajectories (nominal level 0.95).}\label{fig:coverage-3re}
 \end{subfigure}
     \begin{subfigure}{\textwidth}
    \centering
    \resizebox{0.96\linewidth}{!}{\input{figures/ci-width-3re.tex}}
    \caption{Average width of confidence
      intervals for individual trajectories.}\label{fig:ci-width-3re}
     \end{subfigure}
     \begin{subfigure}{\textwidth}
    \centering
    \resizebox{0.96\linewidth}{!}{\input{figures/rmw2-3re.tex}}
    \caption{Root mean squared 2-Wasserstein
      error in population-level process.}\label{fig:rmw2-3re}
     \end{subfigure}
  \caption{Summary of three-component
    simulations, comparing AdaStruMM (\legAdaStruMM), Local-GAM
    (\legLocalGAM), HGAM-GS (\legHGAM), PACE (\legPACE), FACE
    (\legFACE) and bayesFPCA (\legBayesFPCA). Dashed lines show oracle
    benchmarks; in panel (b), the dotted line marks nominal 0.95
    coverage. Local-GAM is shown for $n_i \geq 5$, and its
    confidence-interval summaries are omitted. Points and shaded bands
    are estimates and 95\% confidence intervals for each quantity,
    based on 100 simulation runs. The plots are split by number of
    observations per individual, $n_i$. The $x$-axis is the number of
    individuals, $d$.  In some panels, the $y$-axis is truncated for
    readability.}\label{fig:3re-comparison}
\alttext{Panel (a): AdaStruMM has the lowest non-oracle individual-trajectory error. Panel (b): coverage is close to 0.95. Panel (c): its intervals are the narrowest among the non-oracle methods. Panel (d): it has the lowest non-oracle population-level error, with larger oracle gaps than in the two-component case.}
\end{figure}

The three-component model gives a harder correctly specified test, but
the conclusions are the same as in the two-component case
(Fig.~\ref{fig:3re-comparison}). AdaStruMM again gives the strongest
overall performance, with the lowest individual- and population-level
errors, close-to-nominal confidence-interval coverage and narrow
intervals. The gaps to the oracle benchmarks are
larger than in the two-component case, reflecting the greater complexity
of the shared structure, but remain modest: AdaStruMM stays close to the
individual-level oracle across the settings considered, and is very close
to the population-level oracle for \(n_i\geq 5\). In the sparsest
settings the oracle gaps are larger, but AdaStruMM remains substantially
closer to the oracle benchmarks than the competing general-purpose
methods.

\section{Computational timing results}
\label{sec:simulation_timing}

Figure~\ref{fig:simulation_times} brings together the timing results
for the three simulation cases. AdaStruMM is consistently among the
fastest methods and is substantially faster than HGAM-GS and
bayesFPCA, the methods closest to it in statistical performance. In
the largest two-component setting, with \(d=500\) and \(n_i=10\), the
average fitting times were \(0.2\) minutes for AdaStruMM, \(24\)
minutes for bayesFPCA and \(40\) minutes for HGAM-GS. Absolute times
depend on the implementations and computing environment, so the
comparisons are most informative within each setting.

\begin{figure}
  \begin{subfigure}{\textwidth}
    \centering
    \resizebox{0.96\linewidth}{!}{\input{figures/time-2re.tex}}
    \caption{Two-component model.}\label{fig:time-2re}
  \end{subfigure}
  \begin{subfigure}{\textwidth}
    \centering
    \resizebox{0.96\linewidth}{!}{\input{figures/time-3re.tex}}
    \caption{Three-component model.}\label{fig:time-3re}
  \end{subfigure}
  \begin{subfigure}{\textwidth}
    \centering
    \resizebox{0.96\linewidth}{!}{\input{figures/time-sitar.tex}}
    \caption{Growth-curve model.}\label{fig:time-sitar}
  \end{subfigure}
  \caption{Average model-fitting times
    across the simulation settings. Panels (a)--(c) show the
    two-component, three-component and growth-curve cases,
    respectively. The plots compare AdaStruMM (\legAdaStruMM),
    Local-GAM (\legLocalGAM), HGAM-GS (\legHGAM), PACE
    (\legPACE) and FACE (\legFACE), together with bayesFPCA
    (\legBayesFPCA) in the correctly specified cases and SITAR
    (\legSITAR) in the growth-curve case. Local-GAM and SITAR are
    shown only for \(n_i\geq5\). Points and shaded bands are estimates
    and 95\% confidence intervals based on 100 simulation runs. The
    plots are split by number of observations per individual,
    \(n_i\), and the \(x\)-axis is the number of individuals, \(d\).
    In some panels, the \(y\)-axis is truncated for readability.}\label{fig:simulation_times}
\alttext{Across all three simulation cases, AdaStruMM is among the fastest methods. HGAM-GS and bayesFPCA become much slower as the numbers of individuals and observations increase; PACE is intermediate, while FACE and Local-GAM are relatively fast.}
\end{figure}

\end{document}